\documentclass[a4paper,11pt]{article}
\usepackage{jheppub}
\usepackage{latexsym,amsmath,amsfonts,amssymb}
\usepackage{epsfig,graphics,graphicx,pgf,pstricks}
\usepackage{animate}

\newcommand{\be}{\begin{equation}}
\newcommand{\ee}{\end{equation}}

\newcommand{\ans}{{ans\"atze }}

\newcommand{\im} {\operatorname{Im}}
\newcommand{\re} {\operatorname{Re}}

\newcommand{\vev}[1] {\left<#1\right>}

\newcommand{\lts}{Lefschetz Thimbles }

\title{Optimisation of complex integration contours at higher order}

\author{Francis Bursa$^{(1)}$, Michael Kroyter$^{(2)}$}
\affiliation{$^{(1)}$ School of Physics and Astronomy,\\
University of Glasgow,\\
Kelvin Building, University Avenue, Glasgow,\\
G12 8QQ United Kingdom.
\ \\
$^{(2)}$ Department of Mathematics,\\
Holon Institute of Technology,\\
52 Golomb St., Holon 5810201, Israel.}
\emailAdd{francis.bursa@glasgow.ac.uk}
\emailAdd{michaelkro@hit.ac.il}

\abstract{We continue our study of contour deformation as a practical tool for dealing with the sign problem
using the $d$-dimensional Bose gas with non-zero chemical potential as a toy model.
We derive explicit expressions for
contours up to the second order with respect to a natural small parameter and generalise these contours to an ansatz
for which the evaluation of the Jacobian is fast ($O(1)$).
We examine the behaviour of the various proposed contours as a function of
space-time dimensionality, the chemical potential, and lattice size and geometry
and use the mean phase factor as a measure of the severity of the sign problem.
In turns out that this method leads to a substantial reduction of the sign problem and that it becomes more efficient
as space-time dimensionality is increased.
Correlations among contributions to $\im\vev{S}$ play a key role in determining the mean phase factor
and we examine these correlations in detail.}

\keywords{Lattice Field Theories, The Sign Problem}

\begin{document}

\maketitle

\section{Introduction}
\label{sec:intro}

While perturbative field theory gives some of the most accurate predictions in science,
at strong coupling it loses its efficiency and we are usually forced to use numerical simulations.
The main approach for numerical simulations of field theories, as well as of other, e.g., condensed matter, systems,
is the Monte Carlo method. In this approach the Euclidean $e^{-S}$ factor is interpreted as an unnormalised
probability density and high dimensional integrals are evaluated using importance sampling.
While this method generically works very well, in some cases the action is not real and the naive
interpretation of $e^{-S}$ as a probability density fails. Although this problem can be dealt with using
phase quenching, when the imaginary part of the action becomes large, the expressions that one has to evaluate
fluctuate and an exponentially large number of configurations is needed in order to obtain a reliable result.
This is the sign problem.

Many approaches were proposed for dealing with the sign problem, see for
example~\cite{deForcrand:2002hgr,DElia:2002tig,Gavai:2003mf,Aarts:2009hn,Aarts:2009uq,Aarts:2013lcm,Nagata:2016vkn,Alexandru:2017czx,Lawrence:2020kyw,Pawlowski:2021bbu,Lawrence:2021izu}.
One such approach relies on the use of Lefschetz thimbles~\cite{Cristoforetti:2012su} (see~\cite{Witten:2010cx}
for more on Lefschetz thimbles). Lefschetz thimbles are manifolds of real dimension $N$ that live in a complex
space of dimension $N$, which comes from complexification of the original $N$ real degrees of
freedom of the system\footnote{If the original theory already includes complex degrees of freedom one can complexify separately
their real and imaginary parts. We do that in what follows.}.
Over each thimble the imaginary part of the action is constant, hence the sign problem is avoided when integrating over them.
Also, Lefschetz thimbles form a basis of integration cycles in the complexified space.
Hence, the original integration contour can be deformed using (a multi-dimensional version of) Cauchy’s theorem to a linear
combination of thimbles.
Nonetheless, this method also has some drawbacks. In particular, while computational cost is significantly reduced as compared to
that of a system with a sign problem, it can still be relatively high, generically $O(N^4)$. This issue and others
led to attempts to generalise the thimble method, e.g.~\cite{Alexandru:2015xva,Alexandru:2016lsn}, and to attempts to use contour deformations
that do not rely on thimbles at all~\cite{Mori:2017pne,Mori:2017nwj,Alexandru:2018fqp,Bursa:2018ykf,Kashiwa:2018vxr,Mori:2019tux}.

In particular, in~\cite{Bursa:2018ykf} (henceforth ``paper I'') we studied the one dimensional Bose gas with non-zero chemical potential,
which is often used as a toy model for studying the sign problem~\cite{Aarts:2008wh,Cristoforetti:2012su}.
Here, we attempt to go beyond what was accomplished in paper I and generalise its results in several ways.
In paper I we derived an expression for the contour at first order with respect to an expansion that we defined.
Then, we generalised the obtained expression to \ans and studied the various contours.
Here, we describe the expansion at higher order and generalise the \ans appropriately.
Also, in paper I we studied only the one dimensional case. Here, we examine the same theory in various dimensions and examine the effectiveness
of our approach as a function of dimensionality, with various variables kept fixed.
Finally, in paper I we identified an obstacle towards the efficient ($O(N)$) implementation of our approach coming from the
treatment of boundary terms. We proposed several strategies for handling this issue, all of which had some drawbacks.
Here, we propose another approach and examine its efficiency and its limitations.

The rest of the paper is organised as follows:
In the next section we review the results of paper I and develop the second order of the series expansion for the deformed contour.
We then propose a method for a fast evaluation of the Jacobian and construct \ans that generalise the systematic expansion.
In section~\ref{sec:results} we present simulation results. We perform a thorough examination of our approach, varying different parameters.
Our purpose is to understand the strength of our approach as well as its limitations,
in order to be able to efficiently utilise it in the future for other, more realistic and important systems.
We discuss our results in section~\ref{sec:discussion}.

\section{A perturbative expansion for the contours}
\label{sec:alphaSq}

Here we present and derive our approach. In~\ref{sec:Bose} we recall our toy model, the $d$-dimensional
Bose gas with chemical potential, and define the small parameter $\alpha$, which we use for our expansion in~\ref{sec:expansion}.
We see that at any level of the expansion there are many ways for defining the integration contours.
We limit the search to local expressions for the contours for simplicity as well as since in this case one can hope to obtain
expressions for the Jacobian of the transformation that could be efficiently evaluated.
Simplicity is a key concept for us. We do not attempt to construct the absolutely most efficient contours.
Instead we attempt to derive expressions in a straightforward
way, which could hopefully be generalised also to other systems,
while obtaining large enough reduction of the sign problem that would make the evaluation of observables practical enough.
We return to the evaluation of the Jacobian in~\ref{sec:Jacobian}, where we identify that the periodic boundary conditions
pose a challenge to its efficient evaluation. In paper I we proposed several approaches for dealing with this issue,
none of which was completely satisfactory. Here we propose a refinement of one of the methods of that paper.
We also define an ansatz that generalises the systematic second order expansion, as well as the first order ansatz of paper I.

Throughout the construction, our purpose is not to decrease individual phase factors in the action.
In fact, we find in section~\ref{sec:results} that the expectation value for individual phase factors is larger for our contours
than for the undeformed one, but they cancel out each other.
Hence, we try to put the sign problem on its head: Instead of having large factors that cancel each other
among different configurations, we attempt to have large cancellations of the imaginary part for
any given configuration among different contributions to the phase.
While the systematic expansion is explicitly constructed in order to achieve this goal, \ans have the potential to do even
better in this respect, since they can lead to cancellations of higher order terms, Jacobian contributions to the phase,
and contributions from our special treatment of the boundary, described below.

\subsection{The $d$ dimensional Bose gas with chemical potential}
\label{sec:Bose}

Consider the theory of a Bose gas with non-zero chemical potential on a $d$-dimensional cubic lattice.
After rescaling the fields, the action takes the form,
\begin{equation}
\label{action}
S=\frac{1}{\lambda\alpha^2}\sum_{\vec{r}} \Big(\Phi^*_{\vec{r}}\Phi_{\vec{r}}+(\Phi^*_{\vec{r}}\Phi_{\vec{r}})^2
-\alpha\sum_{\nu=0}^{d-1}\big(\Phi^*_{\vec{r}} \Phi_{\vec{r}+\hat{\nu}} e^{-\mu\delta_{\nu,0}}
   + \Phi^*_{\vec{r}+\hat{\nu}} \Phi_{\vec{r}} e^{\mu \delta_{\nu,0}}\big)\Big)\,,
\end{equation}
where we defined
\begin{equation}
\label{alphaDef}
\alpha\equiv \frac{1}{2d+m^2}\,,
\end{equation}
and $m$ is the mass parameter entering the lattice action before rescaling the fields.
In the rest of the paper we set $\lambda=1$ for simplicity.
We assume periodic boundary conditions with one period, $L$ for the time direction
and one period for all the spatial directions $\tilde L$.
The total number of lattice points in then,
\begin{equation}
\label{Vol}
V=L \tilde L^{d-1}\,.
\end{equation}
We can rewrite the action also in terms of two real fields, $u,v$ related to $\Phi$ in the usual way,
\begin{equation}
\Phi_{\vec{r}}=\frac{u_{\vec{r}}+i v_{\vec{r}}}{\sqrt{2}}\,.
\end{equation}
Without deforming the contour the imaginary part of the action is,
\begin{equation}
\label{imScontour0}
\im(S)=\frac{2\sinh\mu}{\alpha}\sum_{\vec{r}} \im\big(\Phi^*_{\vec{r}} \Phi_{\vec{r}+\hat{0}}\big)
  =\frac{\sinh\mu}{\alpha}\sum_{\vec{r}} \big(u_{\vec{r}}v_{\vec{r}+\hat{0}}-u_{\vec{r}+\hat{0}}v_{\vec{r}}\big)\,.
\end{equation}

We attempt to deal with the sign problem by complexifying the component fields $u$ and $v$ and defining a manifold by specifying
the imaginary parts of $u$ and $v$ as functions of the real parts. Hence, we write,
\begin{equation}
u_{\vec{r}}=x_{\vec{r}} +i y_{\vec{r}}\,,\qquad v_{\vec{r}}=\xi_{\vec{r}} + i \zeta_{\vec{r}}\,.
\end{equation}
In terms of the complex field $\Phi$ we can express the complexification by substituting in the action,
\begin{equation}
\label{substitution}
\Phi_{\vec{r}}\rightarrow \Phi_{\vec{r}}=\phi_{\vec{r}} + i \psi_{\vec{r}}\,,\qquad \Phi^*_{\vec{r}}\rightarrow \bar{\Phi}_{\vec{r}}\equiv\phi^*_{\vec{r}} {\blue + i} \psi^*_{\vec{r}}\,.
\end{equation}
Note that ``complex conjugation does not act on the $i$ in front of the $\psi_{\vec{r}}$'' (in blue).
More accurately, $\bar{\Phi}_{\vec{r}}$ is not the complex conjugate of $\Phi_{\vec{r}}$ and we obtain an extended space by considering
 $\Phi_{\vec{r}}$ and $\bar{\Phi}_{\vec{r}}$ as independent variables.
We can now write,
\begin{equation}
\phi_{\vec{r}} = \frac{x_{\vec{r}}+i \xi_{\vec{r}}}{\sqrt{2}}\,,\qquad \psi_{\vec{r}} = \frac{y_{\vec{r}}+i \zeta_{\vec{r}}}{\sqrt{2}}\,,
\end{equation}
and look for an expression for the contour by specifying $\psi_{\vec{r}}=\psi_{\vec{r}}\big(\{\phi_{\vec{r}\,'}\}\big)$.
This is not the most general form for a contour, but it should be sufficiently general for establishing whether this approach is useful.

With the substitution~(\ref{substitution}), the imaginary part of the action for an arbitrary contour is given by,
\begin{equation}
\begin{aligned}
\label{imSfull}
\im(S)=&\frac{2}{\alpha^2}\re\sum_{\vec{r}}\Big(\phi^*_{\vec{r}}\psi_{\vec{r}}\big(1+2|\phi_{\vec{r}}|^2-2|\psi_{\vec{r}}|^2\big)
  -\alpha\cosh\mu \big(\phi^*_{\vec{r}} \psi_{\vec{r}+\hat{0}} + \psi_{\vec{r}} \phi^*_{\vec{r}+\hat{0}}\big)\\
   &-\alpha\sum_{\nu=1}^{d-1}	\big(\phi^*_{\vec{r}} \psi_{\vec{r}+\hat{\nu}} + \psi_{\vec{r}} \phi^*_{\vec{r}+\hat{\nu}}\big)
   -i\alpha\sinh\mu\big(\phi^*_{\vec{r}} \phi_{\vec{r}+\hat{0}} - \psi^*_{\vec{r}} \psi_{\vec{r}+\hat{0}}\big)
\Big)\,.
\end{aligned}
\end{equation}
Note that exact solutions to this equation exist~\cite{Bursa:2018ykf},
\begin{equation}
\label{bad}
\psi_{\vec{r}}=\pm i \phi_{\vec{r}}\,.
\end{equation}
Moreover, the Jacobian for these solutions is constant, so there is no residual sign problem.
However, for these solutions, not only the imaginary part, but actually the whole action vanishes,
as can be seen by inspecting~(\ref{substitution}): In this case
either $\Phi_{\vec{r}}$ vanishes identically or $\Phi_{\vec{r}}^*$ vanishes identically.
Thus, these solutions do not lead to convergent integrals. This can also be seen by directly inspecting the asymptotic
regions of integration.

\subsection{Defining the expansion}
\label{sec:expansion}

Instead of looking for exact solutions, we attempt to solve for $\psi_{\vec{r}}$ in terms of a power series with respect to $\alpha$,
\begin{equation}
\label{PsiExpansion}
\psi_{\vec{r}}=\sum_{j=1}^\infty \alpha^j \psi_{\vec{r}}^{(j)}\,.
\end{equation}
Inspecting~(\ref{imSfull}) we can expect that for large enough $\mu$ the dominant term contributing to the expansion would be proportional
to $e^\mu\alpha$. Thus, the approach should be useful at least up to values of $\mu$ of the order of,
\begin{equation}
\mu_{\max}\sim -\ln\alpha=\ln(m^2+2d)\,.
\end{equation}
Hence, we can interpret this expansion either as an expansion around $m=\infty$ or as one around $d=\infty$.
For example, while for $d=1$, $m=1$, one can expect to get with the expansion to around $\mu=\ln(3)\simeq 1.1$,
for $d=4$, $m=1$, one could expect to get to about $\mu=\ln(9)\simeq 2.2$.
Note that this is only a rule of thumb. In fact we can get to higher values of $\mu$.

Substituting the series~(\ref{PsiExpansion}) in the expression for the imaginary part of the action~(\ref{imSfull}) we obtain at the lowest order,
\begin{equation}
\label{imS0}
\re\sum_{\vec{r}}\phi^*_{\vec{r}}\Big(\psi_{\vec{r}}^{(1)}d_{\vec{r}}
   -i\sinh\mu \, \phi_{\vec{r}+\hat{0}}\Big)=0\,,
\end{equation}
where we defined,
\begin{equation}
\label{dr}
d_{\vec{r}} \equiv 1+2|\phi_{\vec{r}}|^2=1+x_{\vec{r}}^2+\xi_{\vec{r}}^2\,.
\end{equation}
A simple solution exists even before summation and before taking the real part of the expression,
\begin{equation}
\label{psi0}
\psi_{\vec{r}}^{(1)}=i\sinh\mu\frac{\phi_{\vec{r}+\hat{0}}}{d_{\vec{r}}}\,.
\end{equation}
We refer to this choice as the ``simple first order contour''.
Note that this is not the most general solution one can obtain even before summation. Using the
fact that only the real part should vanish we find that an extra piece can be added to it,
\begin{equation}
\label{psi0Gen}
\psi_{\vec{r}}^{(1)}=i\sinh\mu\frac{\phi_{\vec{r}+\hat{0}}}{d_{\vec{r}}}+i\phi_{\vec{r}}f_{\vec{r}}\,,
\end{equation}
where $f_{\vec{r}}$ is an arbitrary continuous real function. This looks as if we add to our perturbative solution
a component in the direction of the exact (bad) solution~(\ref{bad}). But one can choose $f_{\vec{r}}$
in such a way that the obtained expression is well behaved.
Interesting choices are $f_{\vec{r}}=-\frac{\sinh\mu}{d_{\vec{r}}}$, which for close values of $\phi_{\vec{r}}$
and $\phi_{\vec{r}+\hat{0}}$ cancels the first term in the r.h.s of~(\ref{psi0Gen}) and $f_{\vec{r}}=\frac{\sinh\mu}{d_{\vec{r}}}$,
which leads to cancellation of some of the terms in~(\ref{imSfull}).
We now set $f_{\vec{r}}=0$ for simplicity.
However, we keep in mind that this option exists and use it later on as a starting point for turning the expressions for $\psi_{\vec{r}}$ into a
more general ansatz, with free parameters that can be chosen such that~(\ref{imSfull}) is reduced as much as possible.

We can write the solution~(\ref{psi0}) in terms of components,
\begin{equation}
y_{\vec{r}}^{(1)}=- \sinh\mu\frac{\xi_{\vec{r}+\hat{0}}}{d_{\vec{r}}}\,,\qquad
\zeta_{\vec{r}}^{(1)} = \sinh\mu\frac{x_{\vec{r}+\hat{0}}}{d_{\vec{r}}}\,.
\end{equation}
We notice that the deformation depends only on nearest neighbours in the temporal direction.
Indeed, since the source of the phase in the undeformed case comes from this direction,
this should be the only coordinate relevant at the lowest order.
Note that the solution includes asymptotic regions in which $\psi_{\vec{r}}$ approaches infinity, namely,
regions for which $\phi_{\vec{r}+\hat{0}} \rightarrow \infty$ with bounded $\phi_{\vec{r}}$.
While such regions do not lead to inconsistencies, as long as the integral remains absolutely convergent,
they can still be problematic for the following
reasons:
\begin{enumerate}
\item Contours that go to infinity and back might lead to terms which would mostly cancel each other and hence
      to a mild sign problem similar to the global sign problem that is obtained in the Lefschetz thimble approach.
\item As $|\psi|$ becomes large so does the Jacobian and in particular, the phase of the Jacobian can become large.
      This could lead to a residual sign problem.
\item Our approach is perturbative with respect to $\psi$. Large values of $|\psi|$ can potentially break the validity of the perturbative approach.
\end{enumerate}
In light of these issues, it could be worthwhile to generalise~(\ref{psi0}) to an ansatz for which $\psi$ is always bounded.
We propose such \ans in the next subsection.

At the next order ($\alpha^2$) we obtain,
\begin{equation}
\label{imS1}
\re\sum_{\vec{r}}\Big(\phi^*_{\vec{r}}\psi_{\vec{r}}^{(2)}d_{\vec{r}}
  -\cosh\mu \big(\phi^*_{\vec{r}} \psi_{\vec{r}+\hat{0}}^{(1)} +  \psi_{\vec{r}}^{(1)} \phi^*_{\vec{r}+\hat{0}}\big)
   -\sum_{\nu=1}^{d-1}	\big(\phi^*_{\vec{r}} \psi_{\vec{r}+\hat{\nu}}^{(1)} + \psi_{\vec{r}}^{(1)} \phi^*_{\vec{r}+\hat{\nu}}\big)
\Big)=0\,.
\end{equation}
Substituting~(\ref{psi0}) to this equation while rewriting some terms in light of the fact that only the real
part contributes leads to,
\begin{equation}
\re\sum_{\vec{r}}\phi^*_{\vec{r}}\Big(\psi_{\vec{r}}^{(2)}d_{\vec{r}}
  - i \cosh\mu \, 
	\sinh\mu\frac{\phi_{\vec{r}+2\hat{0}}}{d_{\vec{r}+\hat{0}}}
	 - i s_{\vec{r}} \Big)=0\,,
\end{equation}
where we defined,
\begin{equation}
s_{\vec{r}} \equiv \sinh\mu\sum_{\nu=1}^{d-1}	\Big(\frac{\phi_{\vec{r}+\hat{\nu}+\hat{0}}}{d_{\vec{r}+\hat{\nu}}}+
 \frac{\phi_{\vec{r}-\hat{\nu}+\hat{0}}}{d_{\vec{r}-\hat{\nu}}} \Big)\,.
\end{equation}
We can write a solution for the second order term (which together with~(\ref{psi0}) defines what we call ``the simple second order contour''),
\begin{equation}
\label{secOrd}
\psi_{\vec{r}}^{(2)} = 
  \frac{i}{d_{\vec{r}}} \Big(\cosh\mu 
	\sinh\mu\frac{\phi_{\vec{r}+2\hat{0}}}{d_{\vec{r}+\hat{0}}} + s_{\vec{r}} \Big)\,.
\end{equation}
We see that the expansion in powers of $\alpha$ turns out to be also an expansion in neighbour-distance.
At the second order, there are contributions from terms with distance 2 in the temporal direction and terms with
distance 1 in both the temporal direction and one spatial direction.
The total distance (the sum of distances) of terms that contribute at the second order is at most 2.

Expressing~(\ref{secOrd}) in term of components we obtain,
\begin{equation}
\label{secOrdComponents}
\begin{aligned}
y_{\vec{r}}^{(2)} =&
  -\frac{1}{d_{\vec{r}}} \Big(\cosh\mu 
	\sinh\mu\frac{\xi_{\vec{r}+2\hat{0}}}{d_{\vec{r}+\hat{0}}} + \im(s_{\vec{r}}) \Big)\,,\\
\zeta_{\vec{r}}^{(2)} =&
  \frac{1}{d_{\vec{r}}} \Big(\cosh\mu 
	\sinh\mu\frac{x_{\vec{r}+2\hat{0}}}{d_{\vec{r}+\hat{0}}} + \re(s_{\vec{r}}) \Big)\,.
\end{aligned}
\end{equation}
While the expressions in terms of complex functions are easier to manipulate,
the expressions in terms of components~(\ref{secOrdComponents}) can be useful for performing
simulations, although one can use the complex variables also in the simulations.
Here and below we write expressions in terms of the real variables for the sake of completeness.

If we choose to retain the arbitrary functions $f_{\vec{r}}$ of~(\ref{psi0Gen}) and look for similar expressions
at the second order we obtain,
\begin{equation}
\psi_{\vec{r}}^{(2)} = 
  \frac{i}{d_{\vec{r}}} \Big(\cosh\mu 
	\big(\sinh\mu\frac{\phi_{\vec{r}+2\hat{0}}}{d_{\vec{r}+\hat{0}}}+\phi_{\vec{r}+\hat{0}} (f_{\vec{r}+\hat{0}}-f_{\vec{r}})\big)
	 + s_{\vec{r}} \Big)+i\phi_{\vec{r}}f^{(2)}_{\vec{r}}\,.
\end{equation}
Now the solution depends (at every lattice point $\vec{r}$) on two arbitrary continuous real functions,
$f_{\vec{r}}$ and $f^{(2)}_{\vec{r}}$.
Again, this expression can guide us towards \ans generalising the simple second order contour~(\ref{secOrd}).

\subsection{Fast evaluation of the Jacobian}
\label{sec:Jacobian}

In order to obtain a fast algorithm we attempt to obtain an upper-block-triangular Jacobian matrix.
First, we have to specify the order of lattice points for defining the entries of this matrix.
Most of the obtained expressions depend only on neighbours to the right of the point in the time
direction\footnote{It is possible to rewrite the sums and obtain expressions with mostly left neighbours or expressions
that depend on neighbours from both sides (see paper I for examples).
A main motivation for writing one-sided expressions is to reduce the cost of simulation time by obtaining
an upper-block-triangular Jacobian matrix whose determinant can be efficiently evaluated.}.
Hence, we use a lexicographic ordering in which the most significant weight is that of the time direction.
If it was not for the periodic boundary conditions this would have sufficed for obtaining a matrix
of the desired form at the leading $\alpha$ order.

In paper I we proposed several ways to deal with the problems coming from the periodic boundary conditions.
None of which was completely satisfactory. Here, we propose another approach, which turns out to be successful
in a given range of parameters, but also has some limitations. Let us first recall and discuss the proposals described in paper I:
\begin{itemize}
\item We can use general algorithms for the evaluation of the Jacobian. This would lead to slow simulations\footnote{The
			numerical cost would be $O(N^4)$ per sweep
			with a straightforward algorithm for matrix multiplication, and a little better, but still worse than $O(N^3)$ per sweep,
			with more elaborated algorithms for matrix multiplication	such as Strassen's or
			Coppersmith–Winograd's algorithms~\cite{1969-strassen,CoppersmithW90}.}.
			This can always be achieved and one can manage in this way lattices up to the order of magnitude of
			a hundred points. However, simulations on large lattices become unpractical.
\item The boundary conditions can be changed to Dirichlet boundary conditions, or one could choose to retain the periodic
			boundary conditions, but to avoid modifying the contour at the rightmost lattice point.
			In both cases, a large phase factor would be introduced by this ``last'' point. Moreover, for $d>1$ this is already
			a problem not at a single point, but at a co-dimension one hyper-surface. Thus, a significant sign problem
			would remain in this case, making the approach inapplicable.
\item Block operations can be performed on the Jacobian matrix that bring it to the desired upper-block-triangular form.
			One drawback of this approach is that explicit expressions obtained in this way for $d>1$ at the second order, or
			for \ans generalising it, would be very cumbersome. Another problem is that the evaluation of the expressions obtained this way
			generically has a cost of $O(N^2)$ per sweep. While this is significantly better than the previous proposals, it is still not as good as
			an $O(N)$ algorithm. It was further proposed to evaluate the Jacobians of the small blocks using a particular algorithm that, by changing
			the way they are stored during the run, leads to the desired $O(N)$ behaviour.
			However, it turned out that this algorithm suffers in some cases from numerical instability,
			stemming from the fact that the inverse of some, potentially singular, matrices has to be evaluated.
			Again, the generalisation of this approach to the $d>1$ case would be quite cumbersome.
			Nonetheless, for the simple first order contour~(\ref{psi0}) it was demonstrated in paper I that the instability does not occur.
			Moreover, in this case lattice points with spatial separation do not influence the contour. Hence this method is applicable
			in this case for general $d$. We compare this method to the new method described below in order to evaluate
			the range of validity of the new method.
\end{itemize}
In light of all that it seems that an approach in which the problematic terms are absent from the Jacobian matrix would be desirable.
This can be obtained by rewriting the terms before summation such that each variable would depend
only on variables to its right (in the time direction). We now propose a way to achieve this goal,
for the first order expansion and then for a second order expansion and for the generalising ans\"atze.

\subsubsection{First order}
\label{sec:firstOrderJac}

Consider again the lowest order equation~(\ref{imS0}) with the solution~(\ref{psi0})
everywhere, except on the summands defined on the hyper-surfaces $\vec{r}=(1,\vec{s})$ and $\vec{r}=(L,\vec{s})$.
The remaining equation is now,
\begin{equation}
\re\sum_{\vec{s}}\bigg( \phi^*_{(1,\vec{s})}\Big(\psi_{(1,\vec{s})}^{(1)}d_{(1,\vec{s})}
   -i\sinh\mu \, \phi_{(2,\vec{s})}\Big)
	+ 
	\phi^*_{(L,\vec{s})}\Big(\psi_{(L,\vec{s})}^{(1)}d_{(L,\vec{s})}
   -i\sinh\mu \, \phi_{(1,\vec{s})}\Big)\bigg)=0\,.
\end{equation}
We can rewrite this equation as,
\begin{equation}
\re\sum_{\vec{s}}\bigg(\phi^*_{(1,\vec{s})}\Big(\psi_{(1,\vec{s})}^{(1)}d_{(1,\vec{s})}
   -i\sinh\mu \, \big(\phi_{(2,\vec{s})} -\phi_{(L,\vec{s})}\big)\Big)
	+ \phi^*_{(L,\vec{s})}\psi_{(L,\vec{s})}^{(1)}d_{(L,\vec{s})}\bigg)=0\,.
\end{equation}
Thus, we choose the form~(\ref{psi0}) for all points not on these hyper-surfaces, while on the hyper-surfaces we choose,
\begin{subequations}
\label{psi0Spec}
\begin{align}
\psi_{(1,\vec{s})}^{(1)} &= i\sinh\mu \frac{\phi_{(2,\vec{s})} - \phi_{(L,\vec{s})}}{d_{(1,\vec{s})}}\,,\\
\psi_{(L,\vec{s})}^{(1)} &= 0\,,
\end{align}
\end{subequations}
or in components,
\begin{subequations}
\begin{align}
y_{(1,\vec{s})}^{(1)} &=- \sinh\mu\frac{\xi_{(2,\vec{s})} - \xi_{(L,\vec{s})}}{d_{(1,\vec{s})}}\,,
& \zeta_{(1,\vec{s})}^{(1)} & = \sinh\mu\frac{x_{(2,\vec{s})} - x_{(L,\vec{s})}}{d_{(1,\vec{s})}}\,,\\
y_{(L,\vec{s})}^{(1)} &= 0 \,,& \zeta_{(L,\vec{s})}^{(1)} & = 0\,.
\end{align}
\end{subequations}

Since the Jacobian matrix is now block diagonal the Jacobian is a product of local terms:
\begin{equation}
J=\prod_{\vec{r}}J_{\vec{r}}\,.
\end{equation}
For $\vec{r}=(L,\vec{s})$ we immediately get $J_{\vec{r}}=1$. For the evaluation of $J_{\vec{r}}$ outside this hyper-surface we can
either express everything in terms of the real variables, or work directly with the complex variables using,
\begin{equation}
\label{complexJ}
J_{\vec{r}}=\frac{D(u_{\vec{r}},v_{\vec{r}})}{D(x_{\vec{r}},\xi_{\vec{r}})} =
    \frac{D(u_{\vec{r}},v_{\vec{r}})}{D(\Phi_{\vec{r}},\bar{\Phi}_{\vec{r}})}
		\frac{D(\Phi_{\vec{r}},\bar{\Phi}_{\vec{r}})}{D(\phi_{\vec{r}},\phi_{\vec{r}}^*)}
		\frac{D(\phi_{\vec{r}},\phi_{\vec{r}}^*)}{D(x_{\vec{r}},\xi_{\vec{r}})}\,.
\end{equation}
Recalling that
\begin{equation}
\Phi_{\vec{r}}=\phi_{\vec{r}}+i\psi_{\vec{r}}=\frac{u_{\vec{r}}+iv_{\vec{r}}}{\sqrt{2}}\,,\qquad
\bar{\Phi}_{\vec{r}}=\phi_{\vec{r}}^*+i\psi_{\vec{r}}^*=\frac{u_{\vec{r}}-iv_{\vec{r}}}{\sqrt{2}}\,,
\end{equation}
we can evaluate the first and last Jacobians obtaining,
\begin{equation}
\frac{D(u_{\vec{r}},v_{\vec{r}})}{D(\Phi_{\vec{r}},\bar{\Phi}_{\vec{r}})}=i\,,\qquad
\frac{D(\phi_{\vec{r}},\phi_{\vec{r}}^*)}{D(x_{\vec{r}},\xi_{\vec{r}})}=-i\,.
\end{equation}
Hence, (\ref{complexJ}) reduces to,
\begin{equation}
\label{Jr}
J_{\vec{r}}=\frac{D(\Phi_{\vec{r}},\bar{\Phi}_{\vec{r}})}{D(\phi_{\vec{r}},\phi_{\vec{r}}^*)}
=\left|
\begin{array}{cc}
1+ i\frac{\partial\psi_{\vec{r}}}{\partial\phi_{\vec{r}}} & i\frac{\partial\psi_{\vec{r}}^*}{\partial\phi_{\vec{r}}}\\
i\frac{\partial\psi_{\vec{r}}}{\partial\phi_{\vec{r}}^*} & 1+ i\frac{\partial\psi_{\vec{r}}^*}{\partial\phi_{\vec{r}}^*}
\end{array}
\right|=1+\left|\frac{\partial\psi_{\vec{r}}}{\partial\phi_{\vec{r}}^*} \right|^2-\left|\frac{\partial\psi_{\vec{r}}}{\partial\phi_{\vec{r}}} \right|^2
 +2i\re\Big(\frac{\partial\psi_{\vec{r}}}{\partial\phi_{\vec{r}}}\Big)\,.
\end{equation}
This expression in more general than what we need here and would also be useful for the evaluation of the Jacobians of the \ans that we introduce in what follows.
However, as long as the dependence on $\phi_{\vec{r}}$ comes only from dependence on the real combination $d_{\vec{r}}$~(\ref{dr}), as is the case here,
the second and third terms cancel out and~(\ref{Jr}) is given by
\begin{equation}
J_{\vec{r}}=1+2i\re\Big(\frac{\partial\psi_{\vec{r}}}{\partial \phi_{\vec{r}}}\Big)
=1+4i\re\Big(\phi^*_{\vec{r}}\frac{\partial\psi_{\vec{r}}}{\partial d_{\vec{r}}}\Big)\,.
\end{equation}
Plugging~(\ref{psi0}) in this expression we obtain for $\vec{r}=(t,\vec{s})$ and $1<t<L$,
\begin{equation}
J_{\vec{r}}=1+4i\alpha\sinh\mu\frac{\im\big(\phi_{\vec{r}}^*\phi_{\vec{r}+\hat{0}}\big)}{d_{\vec{r}}^2}
   =1+2i\alpha\sinh\mu\frac{x_{\vec{r}}\xi_{\vec{r}+\hat{0}}-x_{\vec{r}+\hat{0}}\xi_{\vec{r}}}{d_{\vec{r}}^2}\,,
\end{equation}
and for $\vec{r}=(1,\vec{s})$ we similarly obtain,
\begin{equation}
\begin{aligned}
J_{(1,\vec{s})}&= 1+4i\alpha\sinh\mu\frac{\im\big(\phi_{(1,\vec{s})}^*(\phi_{(2,\vec{s})}-\phi_{(L,\vec{s})})\big)}{d_{(1,\vec{s})}^2}=\\
  &=1+2i\alpha\sinh\mu\frac{x_{(1,\vec{s})}\big(\xi_{(2,\vec{s})}-\xi_{(L,\vec{s})}\big)-\big(x_{(2,\vec{s})}-x_{(L,\vec{s})}\big)\xi_{(1,\vec{s})}}{d_{(1,\vec{s})}^2}\,.
\end{aligned}
\end{equation}

\subsubsection{Second order}

For simplicity we illustrate the construction of second order expressions with fast evaluation of the Jacobian for
the one dimensional case.
Since we modified our choice for $\psi_1^{(1)}$ and $\psi_L^{(1)}$ we have to
examine again all terms in~(\ref{imS1}) in which they appear, namely the terms with $r=1$, $L-1$, or $L$.
All terms in the range $1<r<L-1$ vanish upon the substitution of~(\ref{secOrd}).
Also note that the $s_r$ terms, which are the source of additional challenges for a fast evaluation of the Jacobian,
are absent for $d=1$.

The three special terms give,
\begin{equation}
\begin{aligned}
\re\Bigg(& \phi_1^*\Big(\psi^{(2)}_1 d_1 -i \cosh\mu \sinh\mu\frac{\phi_3}{d_2}\Big)-i \cosh\mu \sinh\mu\phi^*_2\frac{\phi_2-\phi_L}{d_1}+\\
   +& \phi_{L-1}^*\psi^{(2)}_{L-1}d_{L-1}-i \cosh\mu \sinh\mu\phi^*_L \frac{\phi_L}{d_{L-1}}+\\
   +& \phi_L^*\Big(\psi^{(2)}_L d_L-i \cosh\mu \sinh\mu \frac{\phi_2-\phi_L}{d_1}\Big)\Bigg)=0\,.
\end{aligned}
\end{equation}
Here, the first line comes from the $r=1$ term, the second is the $r=L-1$ term and the third one is the $r=L$ term.
Simplifying this expression leads to,
\begin{equation}
\label{secondTermBuondary}
\re\!\Bigg(\!\phi_1^*\Big(\psi^{(2)}_1 d_1 -i \cosh\mu \sinh\mu\frac{\phi_3}{d_2}\Big)+ \phi_{L-1}^*\psi^{(2)}_{L-1}d_{L-1}
   + \phi_L^* \psi^{(2)}_L d_L
	+ 2 i \cosh\mu \sinh\mu\phi^*_2\frac{\phi_L}{d_1}
	\!\Bigg)\!=0.
\end{equation}
Note the last term. The presence of $d_1$ in the denominator prevents us from writing $\psi^{(2)}$ that depends
only on components to its right. Thus, we cannot obtain an upper-block-triangular form for the Jacobian.
We can evaluate the efficiency of using the second order expressions by either a slow algorithm, or by
ignoring the last term in~(\ref{secondTermBuondary}).

Ignoring the last term in~(\ref{secondTermBuondary}), we can choose $\psi^{(2)}_1$ to obey the generic equation~(\ref{secOrd}) while setting
\begin{equation}
\psi^{(2)}_{L-1}=\psi^{(2)}_L=0\,.
\end{equation}
This implies that $J_{L-1}$ and $J_L$ do not change as compared to their first order values.
As for the case $1<r<L-1$, let us note that the expression that enters now in the evaluation of the Jacobian is,
\begin{equation}
\label{secondOrder}
\psi_{r}=\alpha\psi_{r}^{(1)} + \alpha^2 \psi_{r}^{(2)} = 
 i\alpha \sinh\mu\frac{\phi_{r+1} + \alpha\cosh\mu \frac{\phi_{r+2}}{d_{r+1}}}{d_r}\,.
\end{equation}
Hence, now we have,
\begin{equation}
\begin{aligned}
J_{r} &= 1+2i\alpha\sinh\mu\frac{x_r\Big(\xi_{r+1}+ \alpha\cosh\mu \frac{\xi_{r+2}}{d_{r+1}}\Big)-\Big(x_{r+1}+ \alpha\cosh\mu \frac{x_{r+2}}{d_{r+1}}\Big)\xi_r}{d_r^2}\\
   &=1+4i\alpha\sinh\mu\frac{\im\Big(\phi_r^*\big(\phi_{r+1} + \alpha\cosh\mu \frac{\phi_{r+2}}{d_{r+1}}\big)\Big)}{d_r^2}\,.
\end{aligned}
\end{equation}
Similarly, we obtain,
\begin{equation}
\begin{aligned}
J_{1} &= 1+2i\alpha\sinh\mu\frac{x_1\Big(\xi_{2} - \xi_L + \alpha\cosh\mu \frac{\xi_{3}}{d_{2}}\Big)-\Big(x_{2} - x_L+ \alpha\cosh\mu \frac{x_{3}}{d_{2}}\Big)\xi_1}{d_1^2}\\
   &=1+4i\alpha\sinh\mu\frac{\im\Big(\phi_r^*\big(\phi_2 - \phi_L + \alpha\cosh\mu \frac{\phi_3}{d_2}\big)\Big)}{d_r^2}\,.
\end{aligned}
\end{equation}

\subsubsection{A more general ansatz}

In paper I we suggested to use an ansatz
that generalises the functional form of the first order expression. Similarly, here we suggest to use an ansatz that generalises the form
of the second order expression. Again, we consider the one dimensional case for simplicity.
We impose the natural $U(1)$ symmetry of complex variables and examine only expressions that can be simulated
efficiently, i.e. expressions whose Jacobian is upper-block-triangular.

The proposed ansatz takes the following form\footnote{Note, that now there are no powers of $\alpha$ in the definition.
Instead, $\alpha$ influences the values of the fit parameters.},
\begin{equation}
\label{ansatz}
\psi_r=\frac{i}{D_r}\bigg(a_1 \phi_r+a_2 \phi_{r+1}
+ \frac{a_3 \phi_r+a_4 \phi_{r+1}+a_5 \phi_{r+2}}{\tilde D_r}\bigg).
\end{equation}
Here, we defined,
\begin{equation}
D_r\equiv 1+b_1 \left|\phi_r\right|^2 + b_2 \left|\phi_{r+1}\right|^2\,,\qquad
 \tilde D_r \equiv 1+b_3 \left|\phi_r\right|^2 + b_4 \left|\phi_{r+1}\right|^2 + b_5 \left|\phi_{r+2}\right|^2\,,
\end{equation}
and the $a_i$ and $b_i$ are ten real parameters subject to the constraint $\forall k, b_k\geq 0$.
The simple second order contour~(\ref{secondOrder}) is obtained by setting in the ansatz,
\begin{equation}
\begin{aligned}
a_1 &=a_3=a_4=0\,,\qquad a_2=\alpha \sinh \mu\,,\qquad a_5=\alpha^2 \sinh\mu \cosh\mu\,,\\
b_2 &=b_3=b_5=0\,,\qquad \ b_1=b_4=2\,.
\end{aligned}
\end{equation}
If instead we set $a_5=0$ we obtain the simple first order contour~(\ref{psi0}).
Again, the expression~(\ref{ansatz}) cannot be used for $r=L$ or $r=L-1$ if we want to obtain an upper-block-triangular Jacobian matrix.
At most, we can set,
\begin{subequations}
\label{ansatzL}
\begin{align}
\psi_L & = \frac{i}{1+b_2 c+b_1 \left|\phi_L\right|^2}\bigg(a_1 \phi_L + \frac{a_3 \phi_L}{1+(b_4+b_5)c+b_3 \left|\phi_L\right|^2}\bigg),\\
\psi_{L-1} & = \frac{i}{D_{L-1}}\bigg(a_1 \phi_{L-1}+a_2 \phi_L +
    \frac{a_3 \phi_{L-1}+a_4 \phi_L}{1+ b_5 c+ b_3 \left|\phi_{L-1}\right|^2+b_4 \left|\phi_L\right|^2}\bigg),
\end{align}
\end{subequations}
where we dropped terms that would have lead to a non-upper-block-triangular Jacobian in the numerator and replaced them by a constant $c$
(to be discussed below) in the denominator.
We can attempt to compensate for the missing terms as we did in the previous subsection.
However, the ansatz we use was not obtained from setting to zero the imaginary part of the action (at some order).
Hence, it is not clear what should be the form of the compensating terms in this case.
In order to choose these terms we pretend that the ansatz~(\ref{ansatz}) was obtained by setting to zero first and second order terms
similar in form to the actual expressions obtained before.
We write $\psi_r=\psi_r^{(1)}+\psi_r^{(2)}$. Then, we pretend that the first order term solves,
\begin{equation}
\re\sum_{\vec{r}}\phi^*_{\vec{r}}\Big(\psi_{\vec{r}}^{(1)}D_{\vec{r}}
   -i\big(a_1 \phi_{\vec{r}}+a_2\phi_{\vec{r}+\hat{0}}\big)\Big)=0\,,
\end{equation}
and the second order term solves,
\begin{equation}
\re\sum_{\vec{r}}\phi^*_{\vec{r}}\Big(\psi_{\vec{r}}^{(2)}D_{\vec{r}}
   -i\frac{a_3 \phi_{\vec{r}}+a_4\phi_{\vec{r}+\hat{0}}+a_5\phi_{\vec{r}+2\hat{0}}}{\tilde D_{\vec{r}}}\Big)=0\,.
\end{equation}
This is a natural generalisation
of the expressions obtained before for the first two orders of the expansion, that would have led
to a solution of the form of the ansatz~(\ref{ansatz}).
Repeating the methods used in the previous subsection and summing the results we get the special values
that include compensating terms,
\begin{subequations}
\label{ansatz1}
\begin{align}
\psi_1 & = \frac{i}{D_1}\Big(a_1 \phi_1+a_2 (\phi_2-\phi_L) + \frac{a_3 \phi_1+a_4 \phi_2+a_5 \phi_3}{\tilde D_1}-\frac{a_5 \phi_{L-1}}{\tilde D_{L-1}}
     -\frac{a_4 \phi_{L}}{\tilde D_{L}}\Big)\,,\\
\psi_2 & = \frac{i}{D_2}\Big(a_1 \phi_2+a_2 \phi_3 + \frac{a_3 \phi_2+a_4 \phi_3+a_5 \phi_4}{\tilde D_2}
            -\frac{a_5 \phi_{L}}{1+b_4 c+ b_3 \left|\phi_L\right|^2+b_5 \left|\phi_2\right|^2}\Big)\,.
\label{psi2ans}
\end{align}
\end{subequations}
While we managed to compensate for all the terms that appear in the numerators in~(\ref{ansatzL}),
we could not do that for the terms in the denominators.
Thus we replaced in the denominators of~(\ref{ansatzL}) and~(\ref{psi2ans}) terms of the form $\left|\phi_r\right|^2$
that would have destroyed the upper-block-triangular form of the Jacobian by a constant $c\geq 0$.
One can decide to set $c=0$, that is, to ignore these contributions.
However, this would lead to denominators which are too small and hence to deformations that are too large.
In order to prevent problems that this can cause, one could prefer to take the limit $c\rightarrow \infty$, which amounts to completely
dropping these terms. This, however, could also result in contours that are quite far from their desired form.
A natural compromise between these two extreme cases would be to choose $c=\vev{\left|\phi_r\right|^2}$.
Alternatively, one can add the constant $c$ to the list of $a$'s and $b$'s defining the ansatz.
This constant is, however, somewhat different, since it does not influence all lattice points.

Using~(\ref{Jr}) for the evaluation of the Jacobian for the ansatz we obtain,
\begin{equation}
\label{JrAns}
J_r=1-p_r^2+2p_r \re(q_r \phi_r^*)+2i\im(q_r \phi_r^*)\,,
\end{equation}
where we defined,
\begin{subequations}
\begin{align}
\label{prqr}
p_r &\equiv \frac{1}{D_r}\Big(a_1 + \frac{a_3}{\tilde{D}_r}\Big)\,,\\
q_r &\equiv \frac{a_1 \phi_r+a_2 \phi_{r+1}}{D_r^2}b_1
+ \frac{a_3 \phi_r+a_4 \phi_{r+1}+a_5 \phi_{r+2}}{D_r \tilde D_r}\Big(\frac{b_1}{D_r}+\frac{b_3}{\tilde{D}_r}\Big)\,.
\end{align}
\end{subequations}
The special cases~(\ref{ansatzL}) and~(\ref{ansatz1}) correspond to special contributions to the Jacobian.
These still take the form~(\ref{JrAns}) only with the following definitions (the expressions for $p_1$ and $p_2$ are the standard ones,
but we write them anyway for completeness),
\begin{subequations}
\begin{align}
p_1 & \equiv \frac{a_1}{D_1} + \frac{a_3}{D_1 \tilde D_1}\,,
\\ \nonumber
q_1 & \equiv \frac{a_1 \phi_1+a_2 (\phi_2-\phi_L)}{D_1^2}b_1
+ \frac{a_3 \phi_1+a_4 \phi_2+a_5 \phi_3}{D_1 \tilde D_1}\Big(\frac{b_1}{D_1}+\frac{b_3}{\tilde{D}_1}\Big)\\
 &\quad -\frac{a_5 \phi_{L-1}}{D_1 \tilde D_{L-1}}\Big(\frac{b_1}{D_1}+\frac{b_5}{\tilde D_{L-1}}\Big)
 - \frac{a_4 \phi_L}{D_1 \tilde D_L}\Big(\frac{b_1}{D_1}+\frac{b_4}{\tilde{D}_L}\Big),
\end{align}
\begin{align}
p_2 & \equiv \frac{a_1}{D_2} + \frac{a_3}{D_2 \tilde D_2}\,,
\\ \nonumber
q_2 & \equiv \frac{a_1 \phi_2+a_2 \phi_3}{D_2^2} b_1
+ \frac{a_3 \phi_2+a_4 \phi_3+a_5 \phi_4}{D_2 \tilde D_2}\Big(\frac{b_1}{D_2}+\frac{b_3}{\tilde{D}_2}\Big)\\
 &\quad -\frac{a_5 \phi_L}{D_2\big(1 +b_4 c+ b_3 \left|\phi_L\right|^2
   + b_5 \left|\phi_2\right|^2\big)}\Big(\frac{b_1}{D_2}+\frac{b_5}{1 +b_4 c+ b_3 \left|\phi_L\right|^2 + b_5 \left|\phi_2\right|^2}\Big),\\
p_{L-1} & \equiv \frac{a_1}{D_{L-1}} +
    \frac{a_3}{D_{L-1}\big(1+ b_5 c+b_3 \left|\phi_{L-1}\right|^2+b_4 \left|\phi_L\right|^2\big)},
\\ \nonumber
q_{L-1} & \equiv \frac{a_1 \phi_{L-1}+a_2 \phi_L}{D_{L-1}^2}b_1 \\
 &\quad + \frac{a_3 \phi_{L-1}+a_4 \phi_L}{D_{L-1}\big(1+ b_5 c+b_3 \left|\phi_{L-1}\right|^2+b_4 \left|\phi_L\right|^2\big)}\cdot\\
 &\quad \nonumber
   \cdot 
		  \Big(\frac{b_1}{D_{L-1}}+\frac{b_3}{1+ b_5 c+b_3 \left|\phi_{L-1}\right|^2+b_4 \left|\phi_L\right|^2}\Big),\\
p_L & \equiv \frac{1}{1+b_2 c+b_1 \left|\phi_L\right|^2}\Big(a_1 + \frac{a_3}{1+(b_4+b_5)c+b_3 \left|\phi_L\right|^2}\Big),
\\ \nonumber
q_L & \equiv \frac{a_1 \phi_L}{\big(1+b_2 c+b_1 \left|\phi_L\right|^2\big)^2}b_1 \\
 &\quad + \frac{a_3 \phi_L}{\big(1+b_2 c+b_1 \left|\phi_L\right|^2\big)\big(1+(b_4+b_5)c+b_3 \left|\phi_L\right|^2\big)}\cdot\\
 &\quad \nonumber
   \cdot \Big(\frac{b_1}{1+b_2 c+b_1 \left|\phi_L\right|^2}+\frac{b_3}{1+(b_4+b_5)c+b_3 \left|\phi_L\right|^2}\Big).
\end{align}
\end{subequations}

\section{Simulation results}
\label{sec:results}

In this section we perform simulations in order to examine the proposed approach identifying both its strengths and its weaknesses.
We concentrate on the mean phase factor as a characteristic of the sign problem.
We examine its behaviour, as a function of $\mu$, in~\ref{sec:varyMu} for the simple first order contour
with the proposed algorithm for treating the special point, as well as using the methods of paper~I, where all
points are treated in the same way.
This is important since while for the first order contour we can use the method of paper~I,
for the more sophisticated contours a numerical instability, described in paper I, prevents us from doing so. Thus,
in the case of this contour, we can disentangle problems stemming from increasing $\mu$ and problems whose source is the special point.
We recognize that at large values of $\mu$ the treatment of the special point reduces the mean phase factor significantly.
In~\ref{sec:origin} we thoroughly examine the origin of this reduction of the mean phase factor.

Next, we examine the dependence of the mean phase factor on lattice geometry in~\ref{sec:geometry} and on
dimensionality in~\ref{sec:fixedAlpha} and in~\ref{sec:fixedM}.
In particular we examine whether the method works better or worse as $d$ is increased when the special point prescription is used,
since two opposite effects exist in this case: On the one hand, our expansion can be interpreted as one around $d=\infty$,
which suggests that as we increase $d$ the behaviour should improve. On the other hand, the special point at $d=1$
becomes, for $d>1$, a co-dimension one hyper-surface, over which the phase accumulates. This suggests that the behaviour should worsen for larger $d$.

Then, in~\ref{sec:DiffContours} we compare the various proposed contours, as a function of $\mu$ and of lattice size.
We find that the second order ansatz behaves significantly better than the first order ansatz of paper I.

Note that for the proposed ans\" atze, large values for the parameters in the numerator can cause a runaway toward wrong asymptotic regions
in the complexified space, which could lead to erroneous results,
unless these parameters are accompanied by large enough values of the matching parameters in the denominator. Hence, we limit the values of the
parameters in the denominator to be not too small.

In all the simulations performed in this work the imaginary part of the phase is consistent with zero,
that is, it is small and within a couple of standard deviations from zero.
Hence, when we mention the mean phase factor we implicitly refer
to its real part. Similar remarks apply to the observables $\vev{S}$ and $\vev{n}$ mentioned below.

Simulations used for determining optimal parameters for the \ans were performed with 50,000 sweeps
on small ($L=8$) lattices and then fine tuned on larger lattices with longer simulation times.
All other simulations were performed with short thermalisation (10,000-30,000 sweeps) followed by 300,000 sweeps.
Running times (on a standard laptop) were approximately linear with lattice size for fixed $d$ and were somewhat longer for larger $d$
and varied from about a quarter of a minute for the short lattices ($d=1$, $L=8$) and up to several hours for the very large ones
($d=1$, $L=7,500$), with several simulations running simultaneously.

\subsection{Varying $\mu$}
\label{sec:varyMu}

\begin{figure}[htb]
\begin{center}
  \includegraphics[width=4.3in]{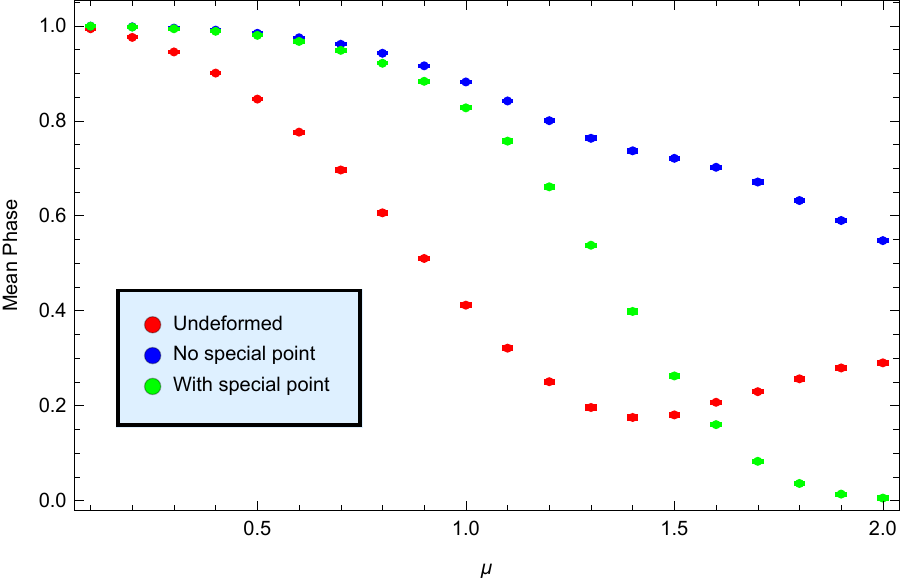}
	\caption{The mean phase factor as a function of $\mu$ for $d=1$, $L=8$, and $m=1$. For small values of $\mu$ the symmetric treatment
					 of all points and the new approach in which there is a special point lead to similar phase factors.
					 From about $\mu=1$ the results begin to diverge, with the phase of the formulation with the special point being inferior
					 and at some stage even worse than that of the undeformed contour.}
\label{fig:comparePhaseD1}
\end{center}
\end{figure}

\begin{figure}[htb]
\begin{center}
  \includegraphics[width=4.3in]{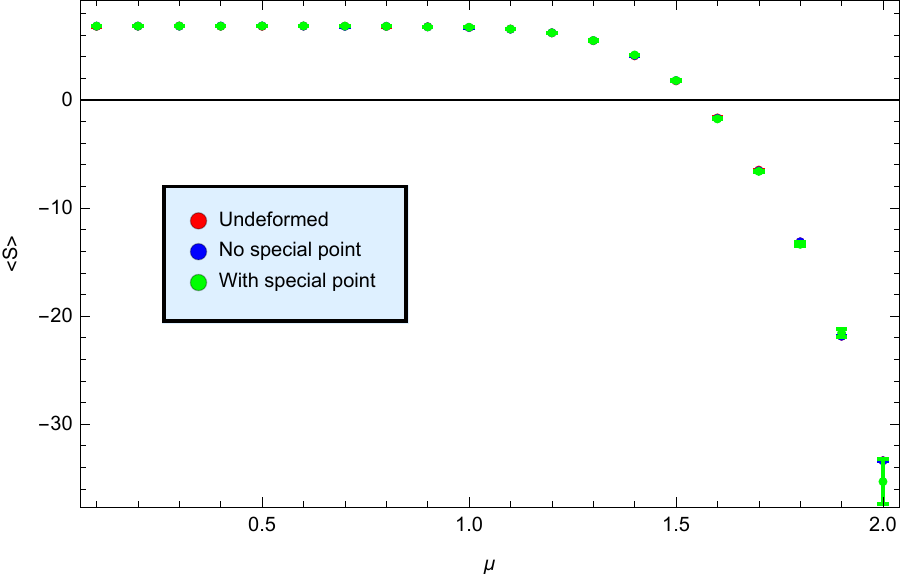}
	\caption{Expectation values of the action for the same three contours. The results are consistent. In fact, for most values of $\mu$
					 one cannot distinguish the different points at the given resolution. This changes only around $\mu=2$, where non-negligible
					 error bars appear, due to the sign problem that is present in this case, for the contour with a special point.}
\label{fig:compareActionD1}
\end{center}
\end{figure}

We proposed a new way to obtain an efficient evaluation of the Jacobian.
However, this method treats one point (for $d=1$) or a specific hyper-surface (for $d>1$) differently.
This can potentially lead to problematic behaviour, especially for large values of $\mu$.
We compared this method to that of paper I for the simple first order contour, for which the method of paper I is applicable
(this contour has no numerical instability problem).
We examine the behaviour of the mean phase factor as a function of $\mu$ for both contours,
as well as for the undeformed contour for comparison,
for the $d=1$, $L=8$, $m=1$ case in fig.~\ref{fig:comparePhaseD1}.
We also examine for these contours an observable, the expectation value of the action, in order to verify that the simulation results
of both methods are consistent, in fig.~\ref{fig:compareActionD1}. The Silver Blaze phenomenon is well demonstrated in this plot.
The fact that the new method does not work well in the current simple case (the simple first order contour at $d=1$) for
too large values of $\mu$ seems to suggest that this is not the best possible method for large $\mu$. However, for
values up to about $\mu=1$, this method, which is relatively simple to use for $d>1$ as well as for the \ans, works well.

\begin{figure}[htb]
\begin{center}
  \includegraphics[width=5.8in]{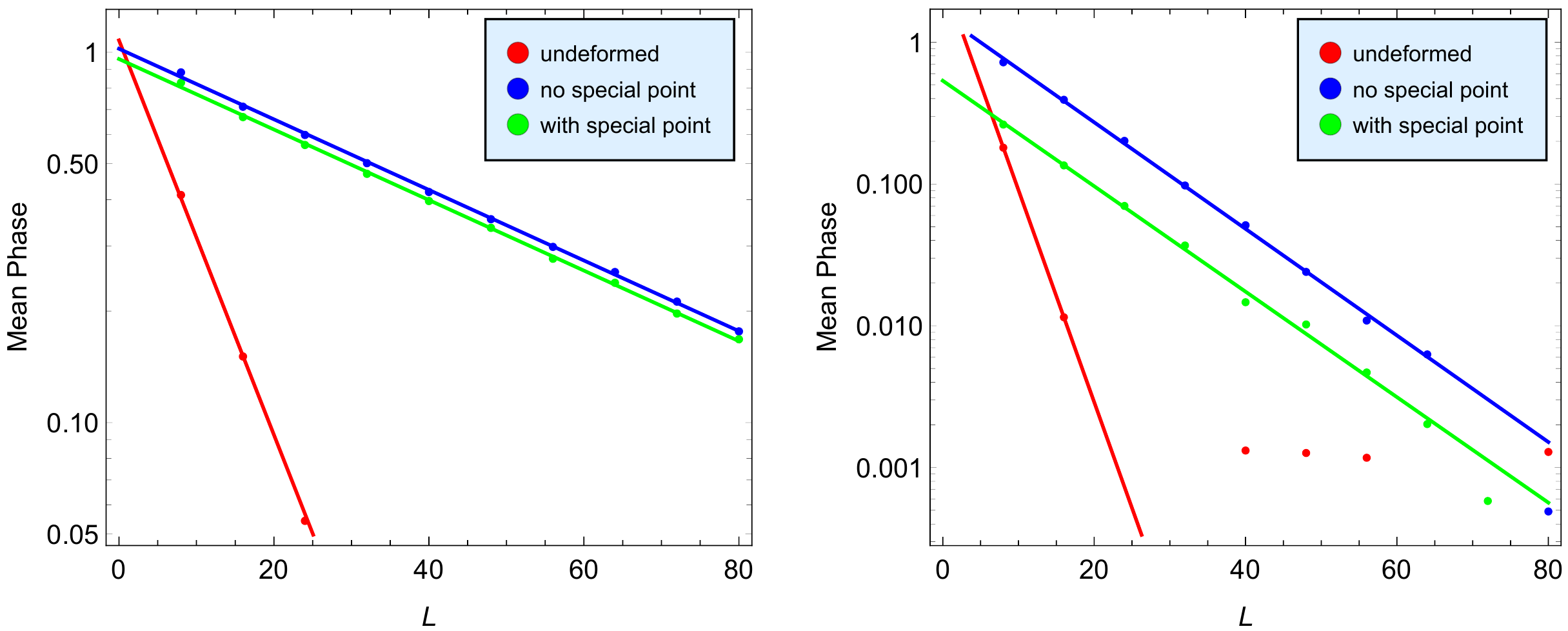}
	\caption{The mean phase factor as a function of lattice size for $\mu=1$ (left) and $\mu=1.5$ (right) on a logarithmic scale.
					 We compare the results of the old algorithm, which treats all point uniformly, the new algorithm that treats one point differently,
					 and the undeformed contour.
					 We observe that the linear fit works very well in all cases, as long as we do not include in the fit points with very low values,
					 i.e., we did not take into account points for which the mean phase factor is below $0.002$, since then
					 the sign problem already becomes significant.
					 We observe that the slopes of the two deformed contours are practically identical (but very different from the undeformed case).}
\label{fig:comparePhaseMuWithL}
\end{center}
\end{figure}

Of course, the values of the mean phase factors for fixed lattice size do not capture the whole story.
We expect this factors to decay exponentially as a function of lattice size, as long as the parameter range with strong sign problem is avoided.
A main factor in the evaluation of the efficiency of a particular contour is this decay rate. We compare the decay rates for $\mu=1$
and $\mu=1.5$ by finding linear fits to the logarithm of the mean phase factors as a function of lattice size in the range $L=8\dots 80$.
The results are shown in fig.~\ref{fig:comparePhaseMuWithL}.
It turns out that in both cases the slopes of the fits are identical within a $1\%$ accuracy.
Thus, if one manages to find an ansatz for which the slope is moderate enough it would be possible to evaluate it using
the new approach up to a large lattice size.

\subsection{The origin of the mean phase factor}
\label{sec:origin}

We want to understand the origin of the difference in the phases between the two approaches.
It is somewhat unexpected, since in both cases the first order term in the expansion of $\im(S)$ exactly vanishes.
To that end we examine contributions to the phase for three cases:
the undeformed contour and the first order ansatz with a particular choice of parameters with and without a special point.
For all these cases we evaluate the rms of the contribution to $\im(S)$ coming from different sites as well as from terms
involving nearest neighbours. We refer to the nearest neighbour pairs as ``half-integer lattice sites'', with the value
being in between the two nearest neighbours involved\footnote{For example, the contribution from a term involving interactions between
lattice sites 4 and 5 is referred to as a contribution from ``lattice point 4.5'' while the contribution from a term involving interactions between
lattice sites 16 and 1 is referred to as a contribution from ``lattice point 16.5''.}.
In all cases $d=1$, $L=16$, and $m=1$. In order to see the effect clearly we work with a large value of $\mu=2$.
We choose for the ansatz $a_1=a_2=0.604$, $b_1=0.9$, $b_2=0.2$ with all other parameters set to zero.
These values are chosen since they lead to a large mean phase factor that decays slowly, for the given choice of $m$ and $\mu$.
For the case with no special points the mean phase factor is 0.72, while with a special point it is 0.099, which is comparable to the
case of the undeformed contour, where it equals 0.051. However, the phase in the case of an ansatz with a special point decays
much slower that that of the undeformed contour, and the decay rate differs from the case with no special point by only about $2.5\%$.
This is similar to the behaviour observed in fig.~\ref{fig:comparePhaseMuWithL}.

\begin{figure}[hbt]
\begin{center}
  \includegraphics[width=3.7in]{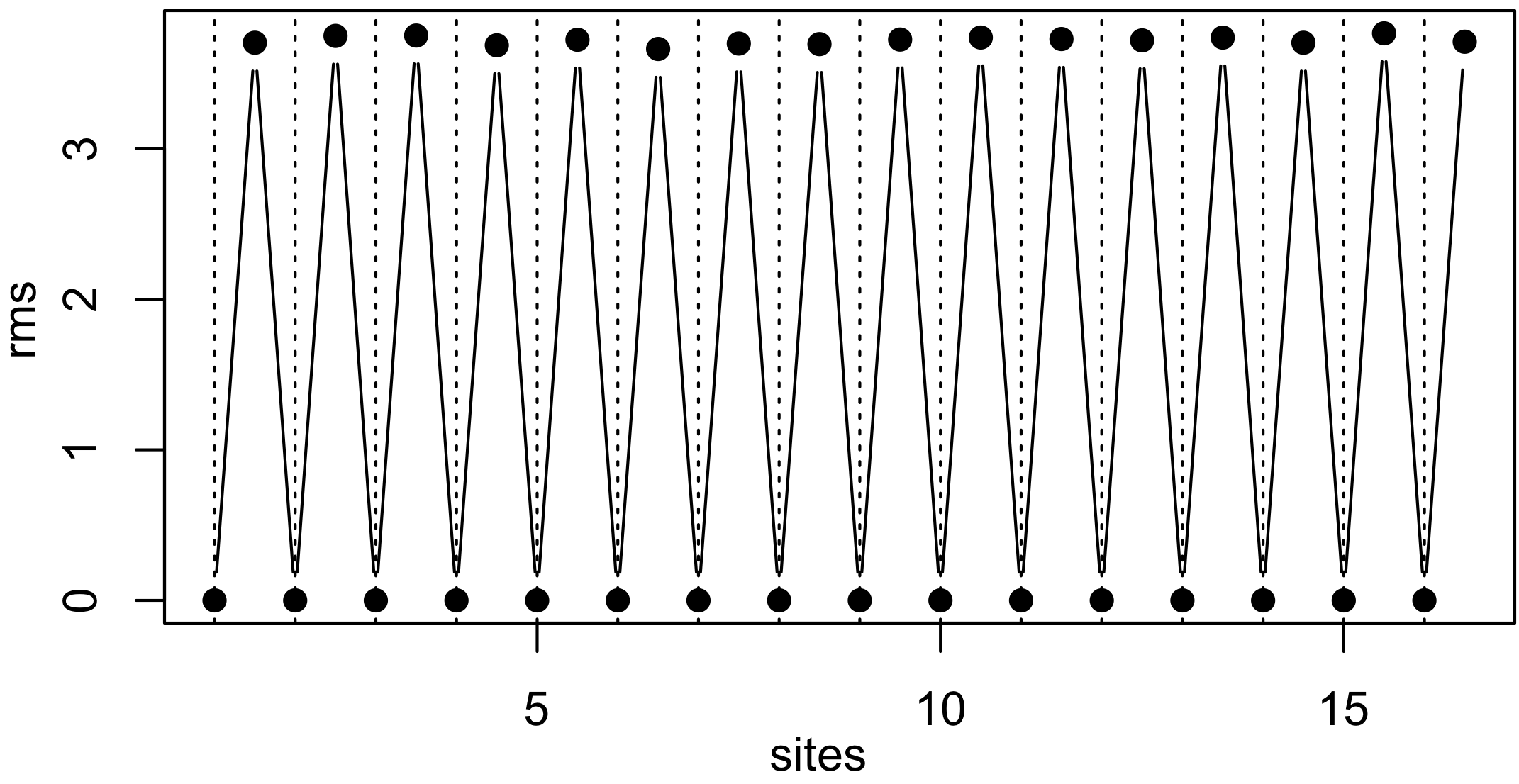}
	\caption{Rms values of contributions to $\im(S)$ as a function of lattice site for the undeformed contour.
					 The contributions from integer lattice sites vanish in this case.}
\label{fig:phaseContributionUndeformed}
\end{center}
\end{figure}

In fig.~\ref{fig:phaseContributionUndeformed} we present the result for the undeformed contour.
In this case the only contributions to $\im(S)$ come from nearest neighbour interactions~(\ref{imScontour0}).
We observe that all 16 contributions are of the same order of magnitude, of about $3.5$. Had all these contributions been
independent we would have obtained a total rms value of $\im(S)$ of the order of $3.5\sqrt{16}=14$.
In fact we observe that the rms value of $\im(S)$ is about $3$, which is significantly lower.
This implies that there are negative 
correlations between these contributions. Indeed, we observe that there are small
and almost uniform negative correlations between contributions from all pairs of ``half-integer lattice sites''.

\begin{figure}[hbt]
\begin{center}
  \includegraphics[width=3.7in]{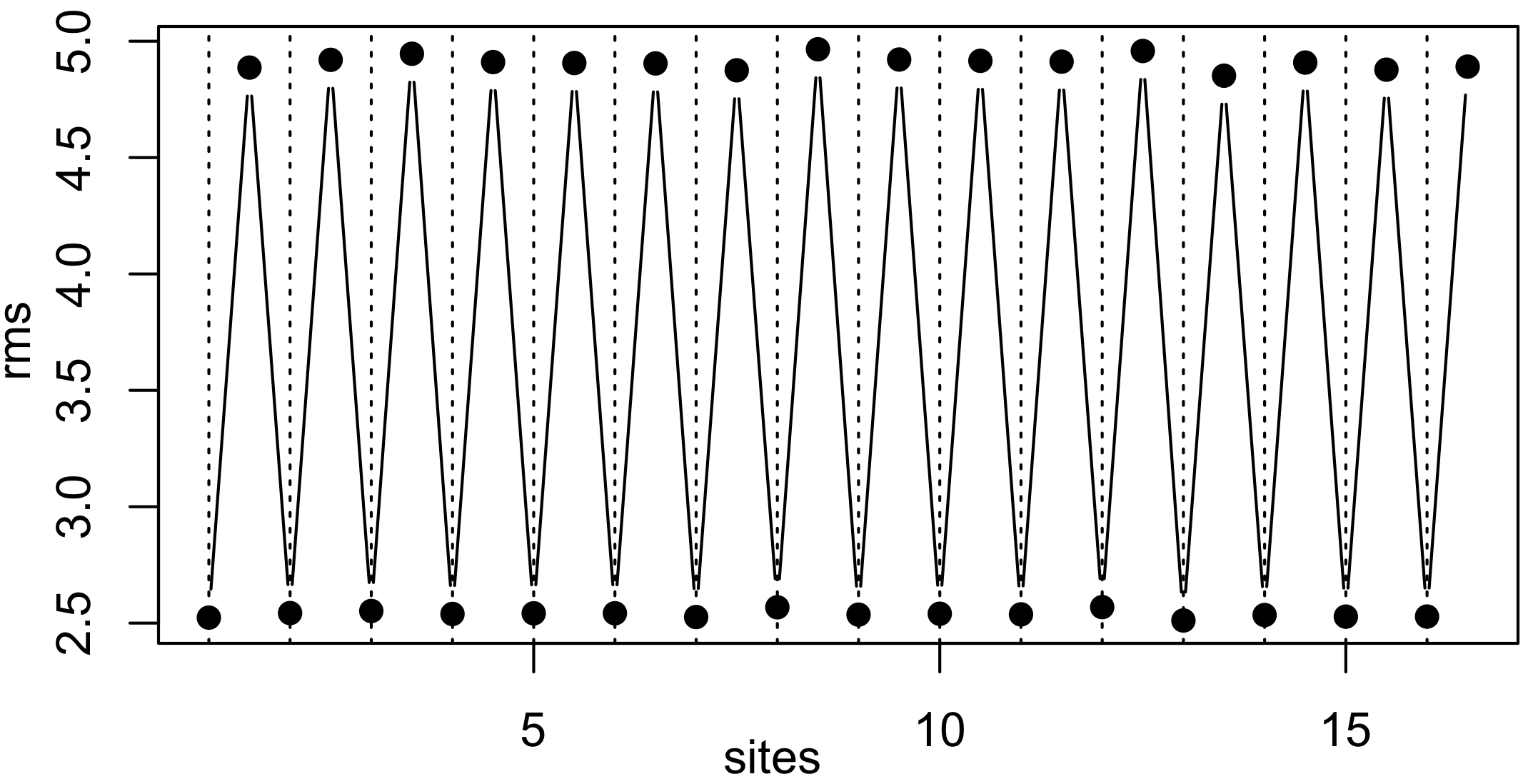}
	\caption{Contributions to $\im(S)$ for the contour corresponding to the ansatz with no special point.
					 The rms values at the (integer) lattice sites is now non-zero, although it is still smaller than the contributions from
					 nearest neighbour pairs (``half-integer lattice sites'').}
\label{fig:phaseContributionAnsatzNoSpecialPt}
\end{center}
\end{figure}

\begin{figure}[hbt]
\begin{center}
  \includegraphics[width=3.7in]{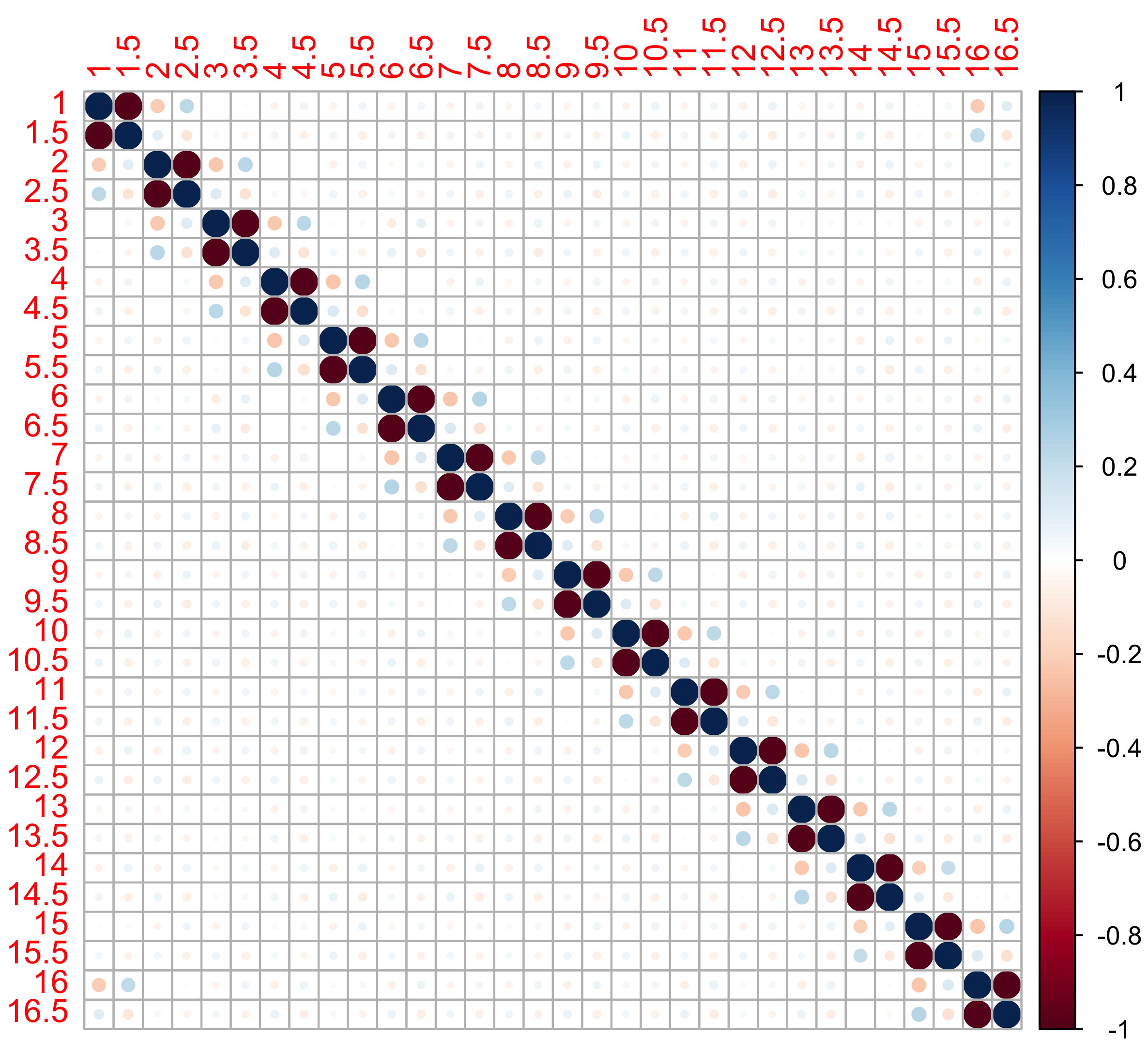}
	\caption{Correlations among contributions to $\im(S)$ from integer and half-integer lattice sites.
					 Positive correlations appear as blue dots and negative ones are depicted as red. The size of the dots represents the size of the correlation.
					 There are small positive and negative correlations almost between all possible pairs, but the most dominant feature
					 (other than the trivial unity correlation on the diagonal) is the strong negative correlation between the phase coming from
					 integer lattice sites $r$ and the half integers to their right, $r+1/2$.
					 The breaking of symmetry between left and right stems from the form of our ansatz~(\ref{ansatz}),
					 which, like the simple first order contour~(\ref{psi0}),
					 includes only dependence of $\psi_r$ on the value of the field to its right $\phi_{r+1}$. These strong anti-correlations are not
					 enough for explaining the phase cancellations, since the size of the contributions at integer and half-integer values differs
					 (recall fig.~\ref{fig:phaseContributionAnsatzNoSpecialPt}).
					 An added effect is the smaller negative correlations between two neighbouring half-integer sites, $r+1/2$ and $r+3/2$,
					 as well as between two neighbouring integer sites, $r$ and $r+1$.
					 This effect is stronger than the small positive correlations between sites $r$ and $r+3/2$ and between $r$ and $r-1/2$, again, in light of
					 the different size of contributions from integer and half-integer sites.}
\label{fig:correlationAnsatzNoSpecialPt}
\end{center}
\end{figure}

Consider now the contour that corresponds to the ansatz with no special point. We present contributions to the rms of $\im(S)$
coming from different lattice sites and from nearest neighbour terms in fig.~\ref{fig:phaseContributionAnsatzNoSpecialPt}.
It is somewhat surprising that the rms contributions from this contour, which has a much higher mean phase factor, are
actually larger than those of the undeformed contour.
This can only happen if there are stronger negative correlations
to balance the large values. We present these correlations in fig.~\ref{fig:correlationAnsatzNoSpecialPt}.
We observe that the phase cancellations result from a delicate and uniform 
correlation among the different contributions to the phase.
One could even claim that a good ansatz is one that maximises the cancellations.

\begin{figure}[hbt]
\begin{center}
  \includegraphics[width=3.7in]{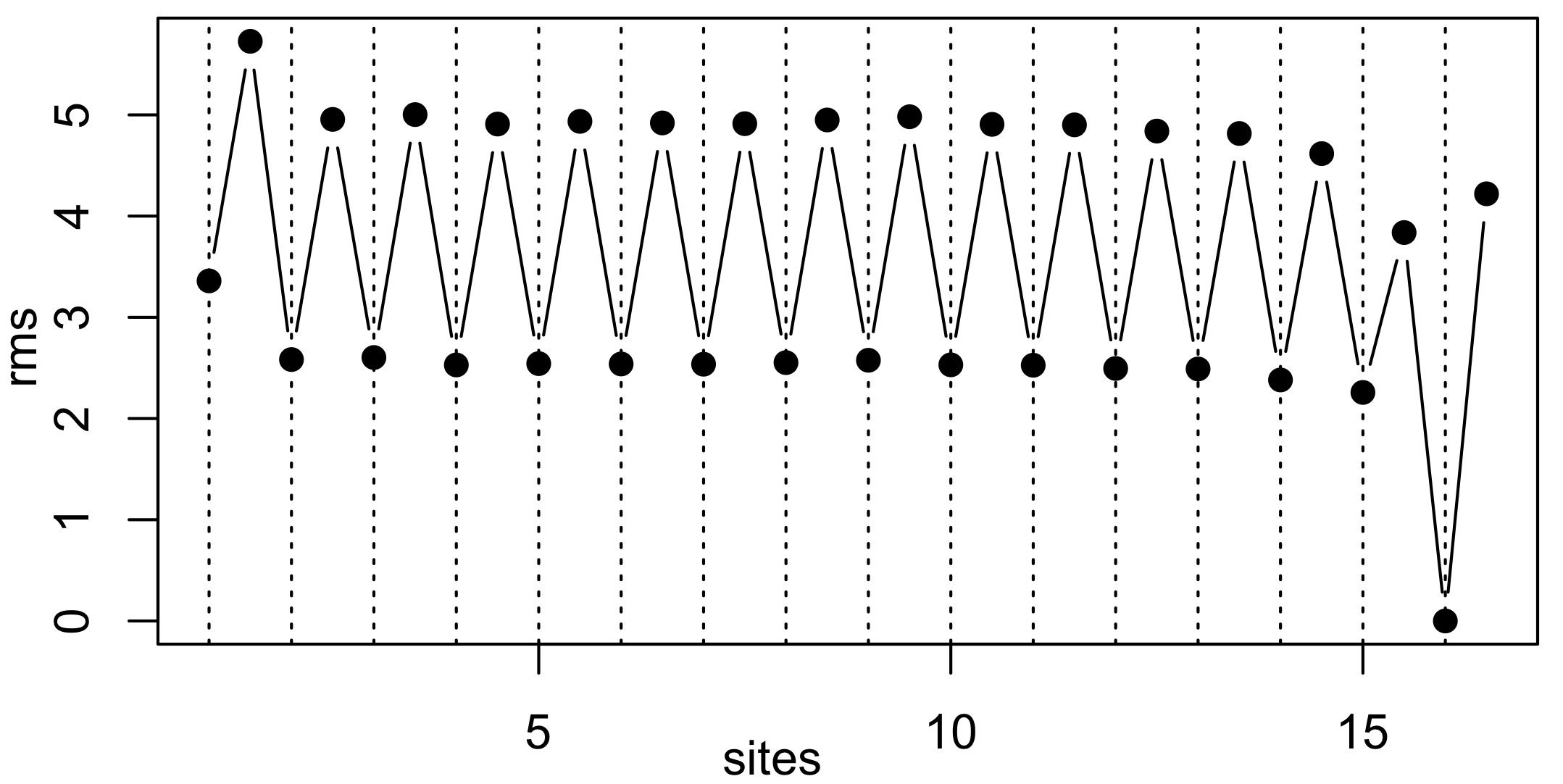}
	\caption{Contributions to $\im(S)$ for the contour corresponding to the ansatz with special point.
					 The values near the special point (from about $14.5$ and up to $1.5$) differ significantly from those of fig.~\ref{fig:phaseContributionAnsatzNoSpecialPt}.}
\label{fig:phaseContributionAnsatzSpecialPt}
\end{center}
\end{figure}

\begin{figure}[hbt]
\begin{center}
  \includegraphics[width=3.7in]{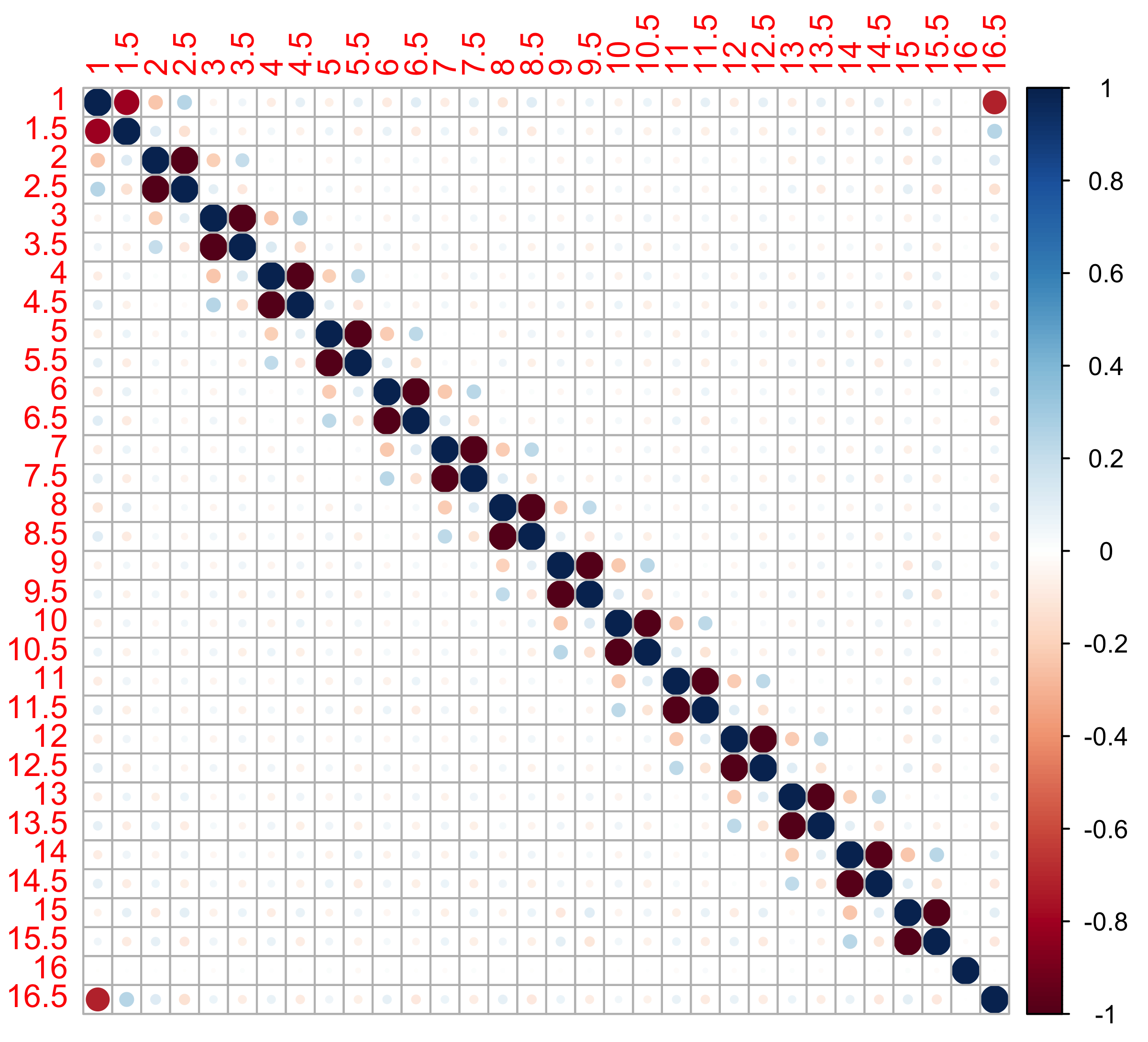}
	\caption{Correlations among contributions to $\im(S)$ from integer and half-integer lattice sites.
					 Positive correlations appear as blue dots and negative ones are depicted as red. The size of the dots represents the size of the correlation.
					 Note that the negative correlation between site 1 and sites $1.5$ and $16.5$ are not as strong as the other ones.}
\label{fig:correlationAnsatzSpecialPt}
\end{center}
\end{figure}

For the same contour with a special point the situation is similar at the bulk but differs near the special point.
The rms contributions to $\im(S)$ are presented in fig.~\ref{fig:phaseContributionAnsatzSpecialPt}, and the correlations
are presented in fig.~\ref{fig:correlationAnsatzSpecialPt}.
The fact that the rms values at a single site are quite large together with the breaking of uniformity of both the rms and the correlations
in some range around the special point are enough to understand the reduction of the phase when a special point is present.
All the observed effects are very sensitive to the value of $\mu$. Hence, for smaller values of $\mu$ the effect is not as dramatic as here.

\subsection{Dependence of the mean phase factor on geometry}
\label{sec:geometry}

For a lattice in $d>1$ of the form defined by~(\ref{Vol}) one can naively expect that the mean phase factor would be independent of geometry,
that is, it would behave in the same way regardless of the choice of $L$ and $\tilde L$, as long as $V$ remain the same.
This expectation relies on the fact that our contours are defined locally and each lattice point has
exactly the same interactions with its neighbours and contributions to $\im(S)$ from these interactions.
In fact there are at least two effects that can modify this expectation.
First, as already mentioned, if we use an algorithm with a special point, there would be an extra contribution
from these points, which is proportional to their number, $\tilde L^{d-1}$. On the other hand,
for small values of $L$, the kinetic term and the periodic boundary conditions would limit the fluctuations
in the temporal directions.

\begin{figure}[hbt]
\begin{center}
  \includegraphics[width=3.7in]{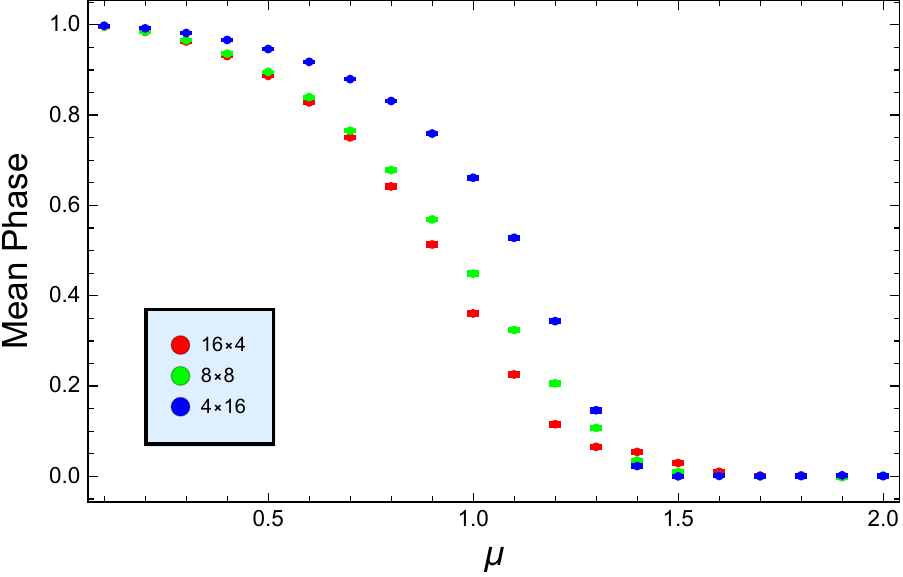}
	\caption{The mean phase factor as a function of $\mu$ for different geometries with 64 lattice points in two dimensions for $m=1$.
					 Different geometries are identified as $L\times \tilde L$.
					 We note that while the $L=4$ case behaves better than the other two cases (for small $\mu$) the difference between the cases with $L=8$
					 and $L=16$ is not very strong. Around $\mu=1.4$ the contribution of the effect of the boundary
					 becomes more significant than that of the effect of the short temporal direction.
					 From this point on small values of $\tilde L$ instead of small values of $L$
					 are associated with better behaviour of the mean phase factor.}
\label{fig:2Dgeometry}
\end{center}
\end{figure}

\begin{figure}[hbt]
\begin{center}
  \includegraphics[width=3.7in]{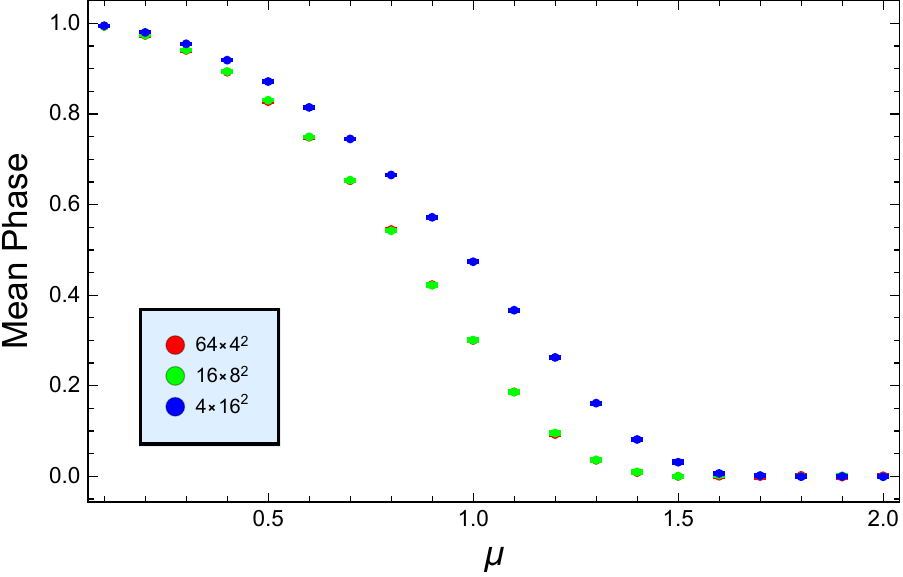}
	\caption{The mean phase factor as a function of $\mu$ for different geometries with 1024 lattice points in three dimensions for $m=2$.
					 Different geometries are identified as $L\times \tilde L^2$.
					 We note that while the $L=4$ case behaves better than the other two cases these other two cases are almost indistinguishable,
					 in line with the naive expectation that it is only the total volume that matters.
					 We can conclude that for $L=16$ the effect of the boundary conditions on uniformization of the field is negligible.
					 The effect of the special point is also not visible. Since the lattice is large, contributions to $\im(S)$ from the bulk
					 reduce the mean phase factor to near zero before this effect becomes important.}
\label{fig:3Dgeometry}
\end{center}
\end{figure}

Since contributions to $\im(S)$ come from these fluctuations~(\ref{action}), short $L$ should result
in larger values of the mean phase factor. One could expect that for small values of $\mu$ the second effect
would be larger while for large values of $\mu$ the first effect would be larger. We examine these expectations for the simple first order contour.
In fig.~\ref{fig:2Dgeometry} we present the results for the two dimensional case and in fig.~\ref{fig:3Dgeometry} for the three dimensional case.
These results are consistent with our expectations.

\subsection{Dependence of the mean phase factor on $d$ for fixed $\alpha$}
\label{sec:fixedAlpha}

An important question is how does the proposed approach depend on space-time dimensionality.
The question by itself is not even well defined since field theories can behave very differently
in different dimensions. But even before examining the continuum limit arises the question:
which parameters should be kept fixed for such a comparison.
A natural possibility for the case at hand is to examine theories with similar values of $\alpha$,
since this is the parameter used in our expansion. Moreover, the terms contributing
to $\im(S)$ depend on $\alpha$ and at the leading order this contribution does not depend on $d$
for fixed lattice size $V$. However, we've already noticed that even a change of the geometry that
does not change the dimensionality can lead to a change in the mean phase factor.
Thus, it is natural to expect that there would be a difference, but of which nature and how
significant would it be?

\begin{figure}[hbt]
\begin{center}
  \includegraphics[width=5in]{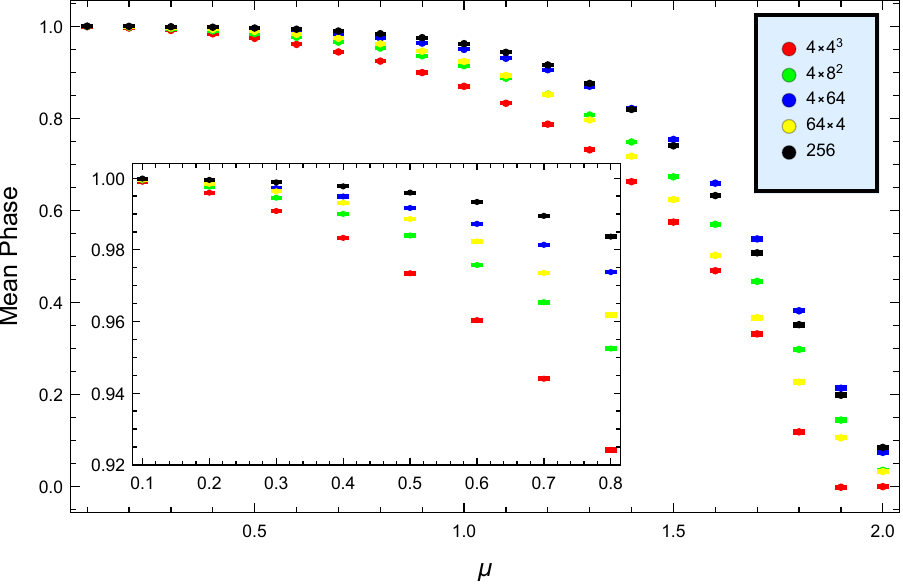}
	\caption{The mean phase factor as a function of $\mu$ for lattices with $256$ sites
					 at various dimensions for $\alpha=\frac{1}{12}$
					 (the mass parameter increases from $m=2$ for $d=4$ to $m=\sqrt{10}\simeq 3.162$ for $d=1$) evaluated using the simple first order contour.
					 The lattices are identified as $L\times\tilde L^{d-1}$. An inset with the range $0.1\leq\mu\leq 0.8$ is presented
					 in order to better distinguish the points in this range.
					 In order to avoid the geometry factor of the previous subsection we compare three lattices with
					 $L=4$, at $d=2,3,4$. We see that indeed the $d=4$ case has a lower value of the mean phase factor than the $d=3$
					 case, whose mean phase factor is still lower than that of the $d=2$ theory.
					 The geometry factor can change this behaviour. Indeed we see that the $d=2$ lattice with $L=64$, whose
					 mean phase factor is consistently lower that that of the $d=2$, $L=4$ case, behaves better than the $d=3$ theory with $L=4$ for small $\mu$,
					 but this changes around $\mu=1.2$.}
\label{fig:D_Comp_alpha}
\end{center}
\end{figure}

The theory at larger value of $d$ is the same as one with lower $d$ with some spatial links added.
While these links do not contribute at the leading order, they still contribute, especially for large values of $\mu$.
These contributions are not accounted for in the simple first order contour.
Thus, at least for this contour, it is expected that larger $d$ would result in a lower
mean phase factor. We examine this expectation in fig.~\ref{fig:D_Comp_alpha}.

\begin{figure}[hbt]
\begin{center}
  \includegraphics[width=5.8in]{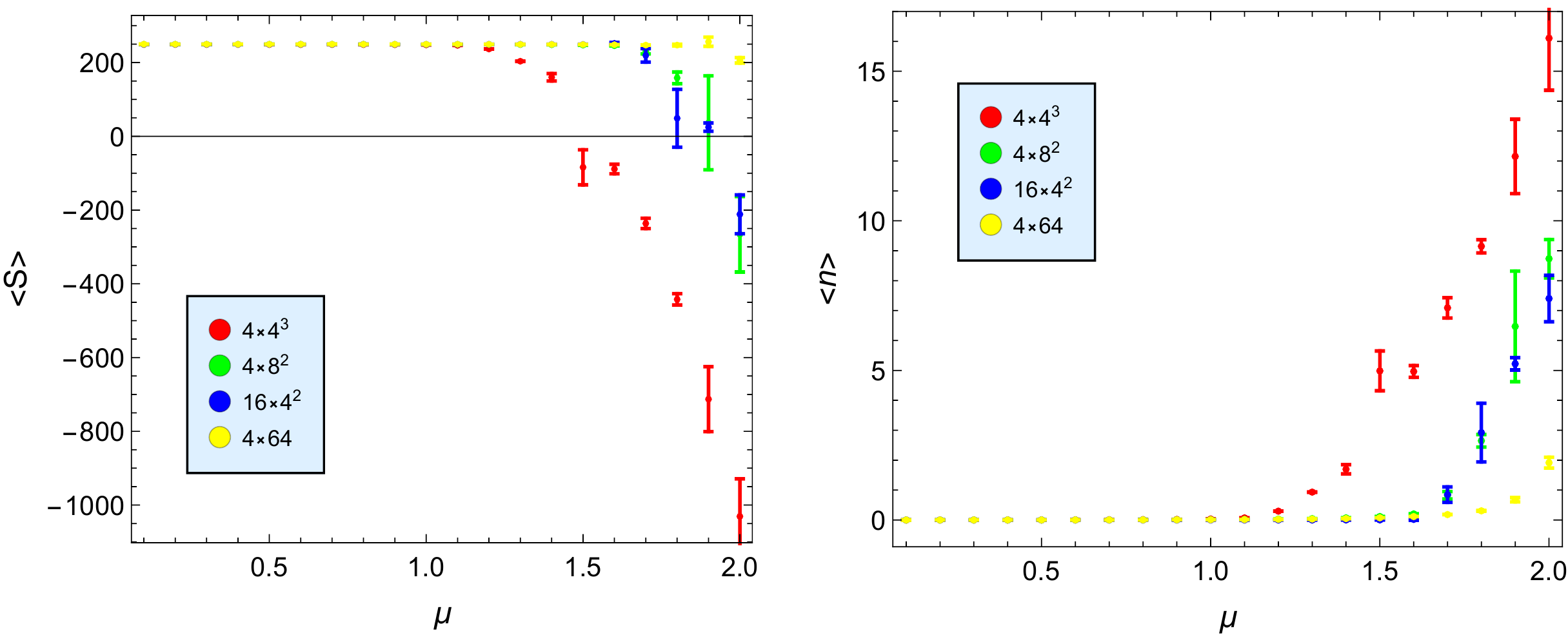}
	\caption{The expectation values of the action (left) and the density (right) as a function of $\mu$ for lattices with $256$ sites
					 at various dimensions for $\alpha=\frac{1}{9}$
					 (the mass parameter increases from $m=1$ for $d=4$ to $m=\sqrt{5}\simeq 2.236$ for $d=2$) evaluated using the simple first order contour.
					 The lattices are identified as $L\times\tilde L^{d-1}$. We observe the Silver Blaze effect starting at different values of $\mu$ for
					 different dimensions. The two lattices at $d=3$ behave similarly; there seems to be no strong dependence on the geometry in this case.
					 As $\mu$ is increased error bars become larger, in light of the reduction of the mean phase factor, which results in a stronger sign problem.}
\label{fig:D_Comp_alpha_ac_dens}
\end{center}
\end{figure}

While it makes sense to compare different dimensions for fixed $\alpha$ from the point of view of the expansion,
observables behave differently in these cases. In fig.~\ref{fig:D_Comp_alpha_ac_dens} we compare two observables,
the action and the density, as a function of $\mu$ for several lattices with fixed $\alpha$. In all cases we
observe the Silver Blaze effect\footnote{We use a different value of $\alpha$ from the one used in fig.~\ref{fig:D_Comp_alpha}. Had we used
the same value we would have obtained almost constant observables due to the Silver Blaze effect.},
but since the values of $m$ differ, the effect begins around different values of $\mu$.

\subsection{Dependence of the mean phase factor on $d$ for fixed $m$}
\label{sec:fixedM}

\begin{figure}[hbt]
\begin{center}
  \includegraphics[width=4.6in]{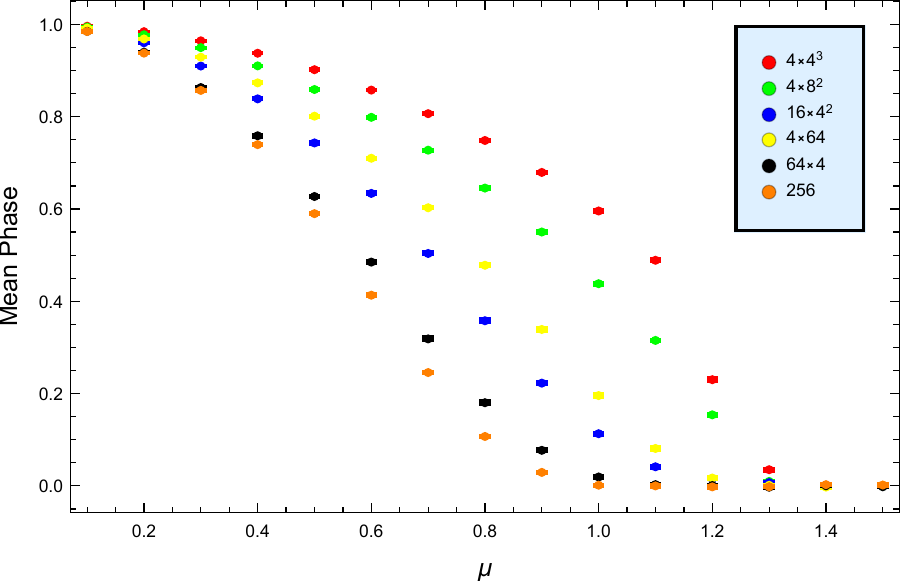}
	\caption{The mean phase factor as a function of $\mu$ for lattices with $256$ sites at various dimensions for $m=1$
					 ($\alpha$ decreases from $\alpha=\frac{1}{3}$ for $d=1$ to $\alpha=\frac{1}{9}$ for $d=4$) evaluated using the simple first order contour.
					 The lattices are identified as $L\times\tilde L^{d-1}$. The expected behaviour of higher mean phase factor for larger $d$ is indeed observed.
					 This can be masked by geometry factors. Indeed, the phase factors of the $4\times 64$ lattice are larger than those of the $16\times 4^2$ one.
					 On the other hand it seems that the effect of the increase of the size of the boundary, where additional contributions to $\im(S)$
					 are present, is not very significant, at least in the current case.}
\label{fig:D_Comp_m}
\end{center}
\end{figure}

\begin{figure}[hbt]
\begin{center}
  \includegraphics[width=5.8in]{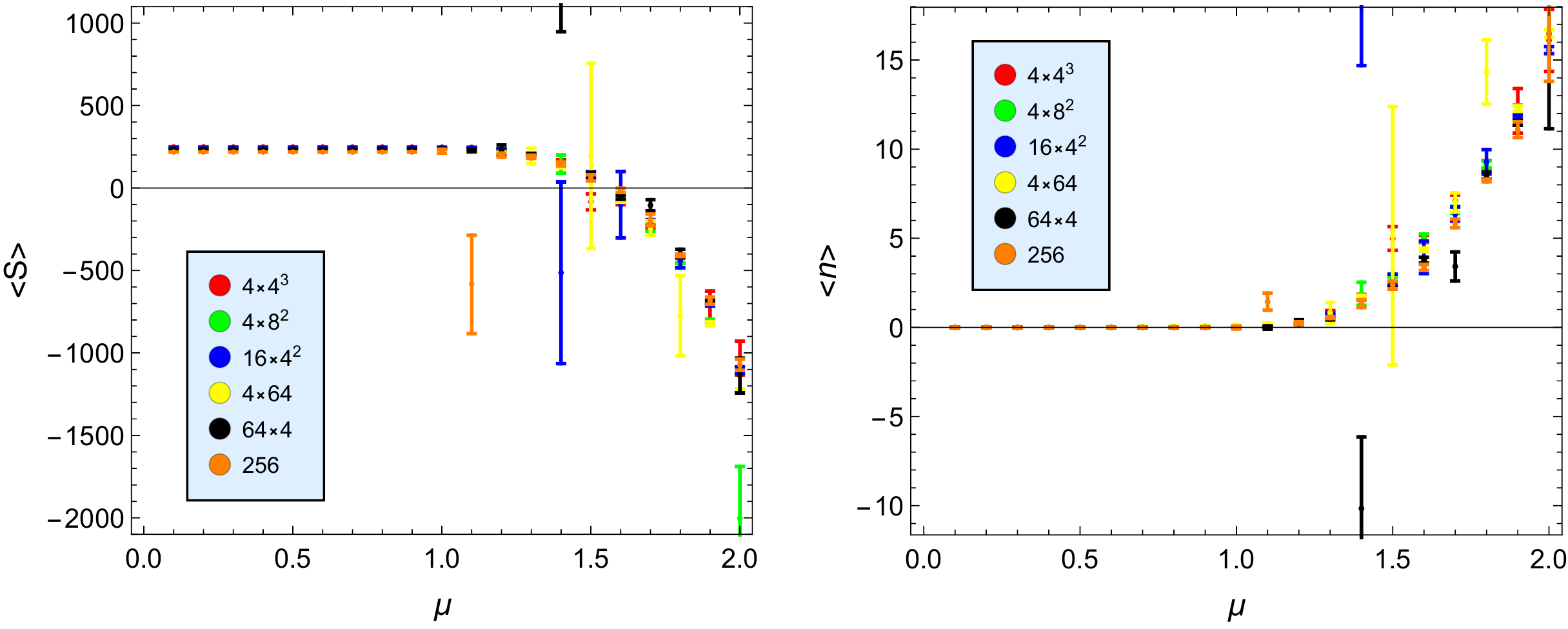}
	\caption{The expectation values of the action (left) and the density (right) as a function of $\mu$ for lattices with $256$ sites
					 at various dimensions for $m=1$ evaluated using the simple first order contour.
					 The lattices are identified as $L\times\tilde L^{d-1}$. The curves almost coincide up to errors coming from the sign problem
					 at large values of $\mu$.}
\label{fig:D_Comp_m_ac_dens}
\end{center}
\end{figure}

As already suggested, a more natural parameter to fix for the comparison of the behaviour at different values of $d$
is the mass parameter $m$. As suggested in section~\ref{sec:alphaSq}, we expect to obtain in this case better behaviour
for larger values of $d$, since from this point of view the expansion can be interpreted as an expansion around $d=\infty$.
We examine this expectation by comparing the mean phase factors in fig.~\ref{fig:D_Comp_m}.
We then examine the Silver Blaze phenomenon in fig.~\ref{fig:D_Comp_m_ac_dens}. Both figures are for $m=1$ and lattices
of 256 sites.

\subsection{Comparing different contours}
\label{sec:DiffContours}

In~\ref{sec:varyMu} we compared the simple first order contour to the undeformed contour and identified that at large $\mu$
the phase is reduced when we use the special point prescription. We defined several other contours, the second order simple contour
and the first and second order ans\"atze. Here we want to compare these contours. Since in some of these cases we have to use the special
point prescription, we use it for all contours, for consistency. We limit the analysis to small and moderate values of $\mu$,
in order to avoid the region in which the special point becomes the dominant variable.
For simplicity we examine the $d=1$ case with $m=1$.

The \ans depend on several parameters and there are several local maxima for the phase in parameter space. We use several different
starting points in parameter space for finding optimal values for the parameters. In particular, we use the parameters that define
the respective simple contours as starting points.
We also use previously obtained values of the parameters at nearby values of $\mu$.
Also, for the second order ansatz we use as starting points the optimal parameters obtained for the first order ansatz.
With all these starting points we probe nearby points in parameter space, searching for parameter values that increase the mean phase factor.
We observe that several different values for the parameters can result in similar local maxima of the mean phase factor.
We are not claiming that the values we use for the parameters are the absolute optimal ones, but they are probably not far from it
and anyway, they are values that we managed to obtain easily, which is what we aim for in this approach.

\begin{figure}[hbt]
\begin{center}
  \includegraphics[width=5in]{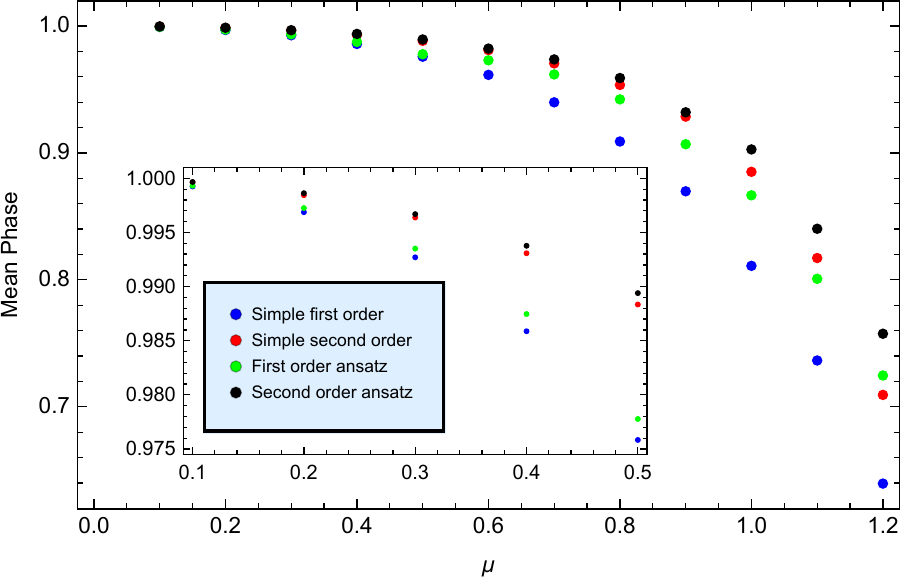}
	\caption{The mean phase factor as a function of $\mu$ for various contours for $m=1$, $L=8$.
					 The values at the range $\mu\leq 0.5$ are inset with a magnified $y$-axis scale for clarity.}
\label{fig:inset}
\end{center}
\end{figure}

In fig.~\ref{fig:inset} we compare the results of the first and second order simple contours and \ans for the $m=1$, $L=8$ case.
We see that for small $\mu$ going to the second order is more important than generalising to an ansatz.
This is not surprising, since this is where we expect the expansion to be particularly useful.
Also, this is where the special point is least important.
As $\mu$ is increased the \ans begin to behave better than the respective simple contours and the gap between
the simple second order contour and the first order ansatz closes. This is again as expected.

\begin{figure}[hbt]
\begin{center}
  \includegraphics[width=4.5in]{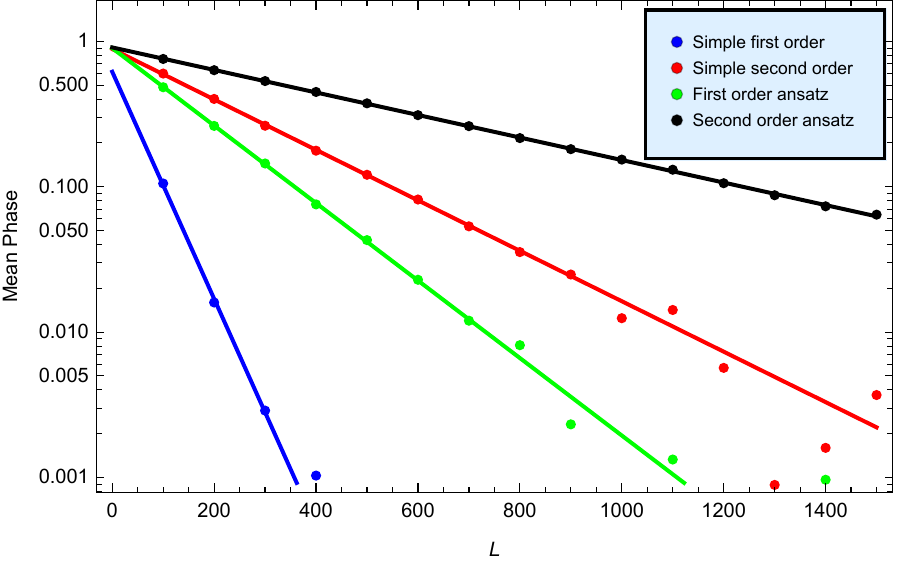}
	\caption{The mean phase factor as a function of lattice size on a logarithmic scale for the various contours for $m=1$, $\mu=1$.
					 The lines are fits to the values at the linear regime. The breaking of the linear (on a logarithmic scale) behaviour
					 is a precursor of the sign problem.}
\label{fig:compareSlopes}
\end{center}
\end{figure}

\begin{figure}[hbt]
\begin{center}
  \includegraphics[width=5.9in]{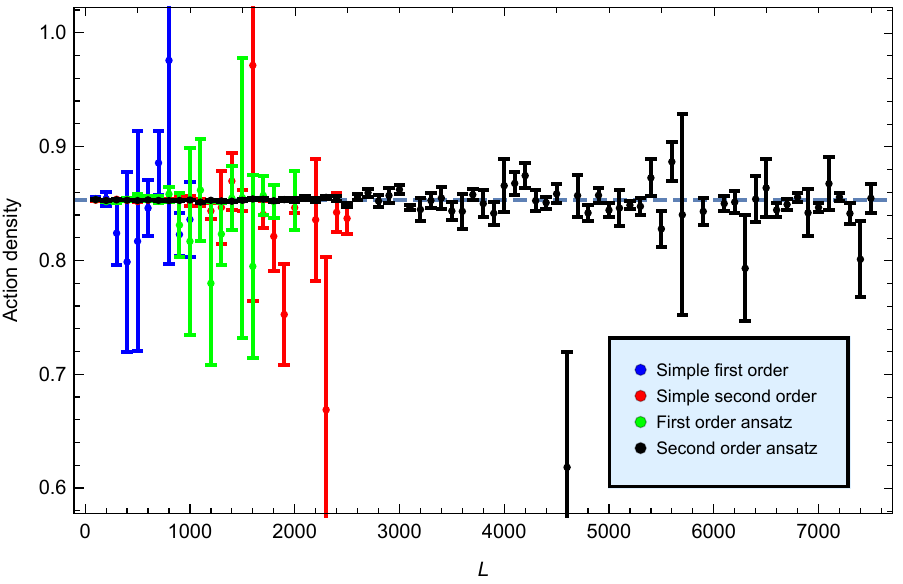}
	\caption{The action density, $\frac{\vev{S}}{L}$, as a function of lattice size for the various contours for $m=1$, $\mu=1$.
					 For each contour we present the results up to somewhere beyond the point where the sign problem begins to become significant.
					 A constant dashed line with the average value of the observable is presented for convenience.
					 For the first order simple contour and the first order ansatz the results are consistent with those of paper I, which
					 implies that the treatment of the special point is of no significant importance.}
\label{fig:compareActionL}
\end{center}
\end{figure}

Nonetheless, it might seem that the benefit from using the simple second order contour or the \ans as compared to the simple first order contour
is not significant.
This impression is incorrect, since what actually matters is the decay rate of the phase as a function of lattice size,
as we examined in fig.~\ref{fig:comparePhaseMuWithL}. We repeat the same analysis for the four contours in the problem at hand.
The results are shown in fig.~\ref{fig:compareSlopes}.
In paper I we claimed that the first order ansatz is reliable up to about $L=1000$. This is consistent with the current result.
We note that the current treatment of the special point hardly changes anything in this respect.
Somewhat surprisingly, the second order simple contour is doing better than the first order ansatz even at this, not too small,
value of $\mu$. Not surprisingly, the second order ansatz behaves even better. We can expect that it would give reliable
results at least up to $L=3000$, that is, by generalising the first order ansatz we can triple the size of the lattice,
which can be used.
In order to examine this expectation we plot the action density, $\frac{\vev{S}}{L}$,
as a function of lattice size, retaining $m=1$ and choosing $\mu=1$, for the various contours, in fig.~\ref{fig:compareActionL}.
For these sizes of the lattice it is expected that this observable is $L$-independent.
We observe that all four contours and especially the second order ansatz give quite reliable results even beyond the point
where the sign problem is expected to become significant.

\section{Discussion}
\label{sec:discussion}

We studied the method of contour deformation and established that it is a viable option for dealing with the sign problem.
Indeed, we managed to get reliable results for quite large lattices for generic values of the parameters.
To that end we used both a systematic expansion and \ans generalising this expansion.
The expansion used can be regarded as one around $m=\infty$, around $d=\infty$, or as an expansion with respect to neighbour
distance. Each one of these interpretations can be used for the construction of generalisations of our approach to other theories.

Both the expansion and the \ans have a large degree of arbitrariness. Our purpose was not to construct an elaborate mathematical
framework, but to establish a practical tool for dealing with the sign problem. From this perspective the arbitrariness is
not a problem but a bliss, since one can deform the contour in many ways and choose different values for the
parameters and they would all be good enough for obtaining the desired results. Thus, it should be relatively easy to
use this method in practice and we believe that similar results would be obtained by employing this approach also to other systems.

One guiding principle we used that restricted this arbitrariness is the requirement of obtaining a computationally efficient
algorithm. This led to a restriction on the form of the contour deformation to one that would lead to Jacobians that are simple to evaluate.
As in paper I, a stumbling block towards an upper-block-triangular
form for the Jacobian matrix originated from the periodic boundary conditions.
We proposed an approach that improved one of the options that we considered in paper I, of not deforming the contour related to
the special point at all. This new approach turned out to be quite adequate for not very large
values of the chemical potential. However, for large values of $\mu$ the contributions to the phase from the special point
become significant. This contribution becomes more problematic in higher dimensions, although generally they behave better.
We identified that, especially for ans\"atze, the origin of this problem is the non-uniformity of correlations that stems from
treating a specific point (or hyper-surface) as special.
Thus, it would be advisable to devise contours that do not treat any point in a special way. We currently examine such possibilities.

Another important direction is the examination of the efficiency of the contour deformation approach for fermionic theories.
The local nature of deformations, which we imposed in order to obtain simple and computationally efficient expressions,
as well as the general requirement of computational efficiency are challenged in this case by the fermionic determinant.
Moreover, it was argued in~\cite{Lawrence:2021izu} that in many models with fermionic sign problems contour deformation would be ineffective.
While there are known examples of contour deformations, e.g., \lts that improve sign problems in fermionic cases, it would be
interesting to examine the current method in several such cases. We hope to study this issue in the future.

\section*{Acknowledgements}

We would like to thank Gert Aarts, Naomi Don-Yechiya, Kouji Kashiwa, Scott Lawrence, Yuto Mori, and Akira Ohnishi for discussions.
The research of M.~K. was supported by the Israel Science Foundation (ISF), grant No. 244/17.


\bibliography{bib}

\vfill\eject

\end{document}